\documentclass[amsmath,amssymb,prl,superscriptaddress,notitlepage]{revtex4-1}
\usepackage[dvipdfmx]{graphicx}
\usepackage{dcolumn}
\usepackage{bm}
\usepackage{subfigure}
\usepackage{booktabs}
\usepackage{color}

\newcounter{one}
\def\ket#1{\mbox{\boldmath $#1$}}
\newcommand{\bracket}[1]{\left\langle #1 \right\rangle}

\newcommand{\cav}[2]{\backslash (#1,#2)}
\newcommand{\affA}{
Artificial Intelligence Research Center, 
National Institute of Advanced Industrial Science and Technology, 
2-3-26 Aomi, Koto-ku, Tokyo, Japan
}
\newcommand{\affB}{
Department of Mathematical and Computing Science,
Tokyo Institute of Technology, 4259-G5-22, Nagatsuta-cho, Midori-ku, Yokohama, Kanagawa 226-8502, Japan
}
\begin{document}
\title{\textbf{Cross-validation estimate of the number of clusters in a network}}
\author{Tatsuro Kawamoto$^{\ast}$}
\affiliation{\affA}
\author{Yoshiyuki Kabashima}
\affiliation{\affB}
\date{\today}
\begin{abstract}
Network science investigates methodologies that summarise relational data to obtain better interpretability. 
Identifying modular structures is a fundamental task, and assessment of the coarse-grain level is its crucial step. 
Here, we propose principled, scalable, and widely applicable assessment criteria to determine the number of clusters in modular networks based on the leave-one-out cross-validation estimate of the edge prediction error. 

\end{abstract}
 
\maketitle

\section*{Introduction}
Mathematical tools for graph or network analysis have wide applicability in various disciplines of science.  
In fact, many datasets, e.g., biological, information, and social datasets, represent interactions or relations among elements and have been successfully studied as networks \cite{Barabasi201608,newman2010networks} using the approaches of machine learning, computer science, and statistical physics. 
In a broad sense, a major goal is to identify macroscopic structures, including temporal structures, that are hidden in the data. 
To accomplish this goal, for example, degree sequences, community and core--periphery structures, and various centralities have been extensively studied. 
Here, we focus on identifying modular structures, namely, graph clustering \cite{Fortunato201075,Leger2013}. 
Popular modular structures are the community structure (assortative structure) and disassortative structure \cite{GirvanNewman2002,Radicchi2004,Newman2006PRE,LFRbenchmark2008}, although any structure that has a macroscopic law of connectivity can be regarded as a modular structure. 
The Bayesian approach using the stochastic block model \cite{holland1983stochastic}, which we will describe later, is a powerful tool for graph clustering. 
In general, graph clustering consists of two steps: 
selecting the number of clusters and determining the cluster assignment of each vertex. 
These steps can be performed repeatedly. 
Some methods require the number of clusters to be an input, whereas other methods determine it automatically. 
The former step is called model selection in statistical frameworks, and this step is our major focus.

\begin{figure*}[t]
 \begin{center}
    \includegraphics[width=0.99 \columnwidth]{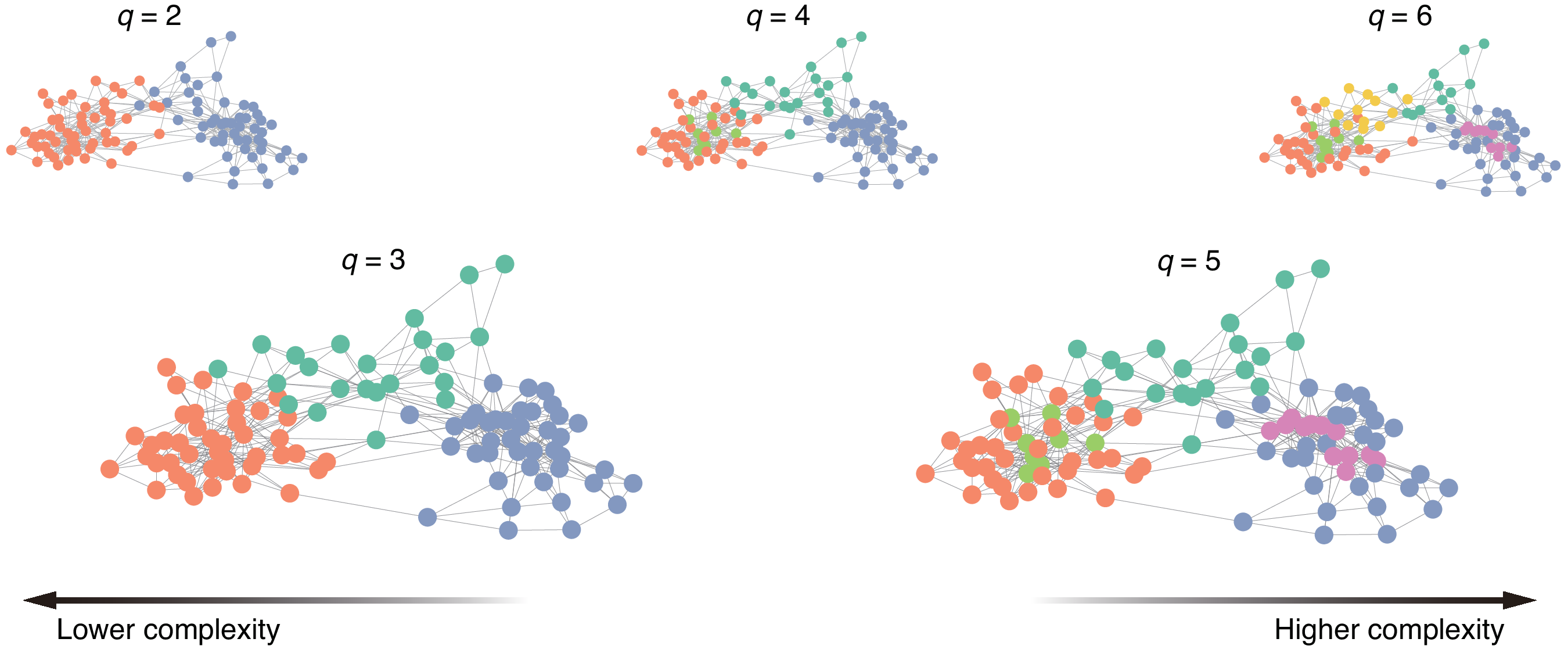}
 \end{center}
 \caption{
 	\textbf{Partitions of the network of books about US politics.} 
	The inferred cluster assignments for various input numbers of clusters $q$. 
	The vertices with the same colour belong to the same cluster. 
	The partition with large $q$ has a higher resolution and is obtained by fitting a model that has a higher complexity. 
}
 \label{fig-politicalbooks-complexity}
\end{figure*}

The selection of the number of clusters is not an obvious task. 
For example, as shown in Fig.~\ref{fig-politicalbooks-complexity}, more complex structures can be resolved by using a model that allows a larger number of clusters. 
However, because the whole purpose of graph clustering is to coarse grain the graph, the partition with excessively high resolution is not desirable. 
This issue is also related to the level of resolution that is required in practice. 
The role of the model selection frameworks and criteria is to suggest a possible candidate, or candidates, for the numbers of clusters to be analysed. 

Which framework and criterion to use for model selection and its assessment is an inevitable problem that all practitioners face, as multiple candidates have been proposed in the literature. 
A classical prescription is to optimise an objective function that defines the strength of a modular structure, such as the modularity \cite{NewmanGirvan2004,ZhangMoore2014} and the map equation \cite{Rosvall2008,Rosvall2011}; combinatorial optimisation among all of the possible partitions yields both the number of clusters and the cluster assignments. 
Although it might seem to be unrelated to statistical frameworks, it should be noted that modularity maximisation has been shown to be equivalent to likelihood maximisation of a certain type of stochastic block model \cite{Newman2013,ZhangMoore2014}. 
One can also use the spectral method and count the number of eigenvalues outside of the spectral band \cite{ShiMalik2000,Luxburg2007,Krzakala2013}; the reason for this prescription is the following: while the spectral band stems from the random nature of the network, the isolated eigenvalues and corresponding eigenvectors possess distinct structural information about the network. 
As another agnostic approach, an algorithm \cite{AbbeNIPS2015} that has a theoretical guarantee of the consistency for the stochastic block model in sparse regime with sufficiently large average degree was recently proposed; a network is called sparse when its average degree is kept constant as the size of the network grows to infinity, and typically, it is locally tree-like. 
Finally, in the Bayesian framework, a commonly used principle is to select a model that maximises the model's posterior probability \cite{Nowicki2001,daudin08,Decelle2011a,Hayashi2016,NewmanReinert2016} or the model that have the minimum description length \cite{PeixotoPRL2013,PeixotoPRX2014,PeixotoPRX2015}, which leads to Bayesian Information Criterion (BIC)-like criteria.

Minimisation of the prediction error is also a well-accepted principle for model selection, and cross-validation estimates it adequately \cite{HastieTibshiraniFriedman,Arlot2010}. 
This approach has been applied to a number of statistical models, including those models that have hidden variables \cite{Celeux2008,Vehtari2012}. 
Although the cross-validation model assessments for the stochastic block model are also considered in \cite{HoffNIPS2007,Chen2014,AiroldiNIPS2008}, they are either not of the Bayesian framework or performed by a brute-force approach. 
A notable advantage of performing model selection using prediction error is that the assumed model is not required to be consistent with the generative model of the actual data, whereas the penalty term in the BIC is derived by explicitly using the assumption of model consistency. 
In this study, we propose an efficient cross-validation model assessment in the Bayesian framework using the belief propagation (BP)-based algorithm.

\begin{figure}[t]
 \begin{center}
   \includegraphics[width=0.4 \columnwidth]{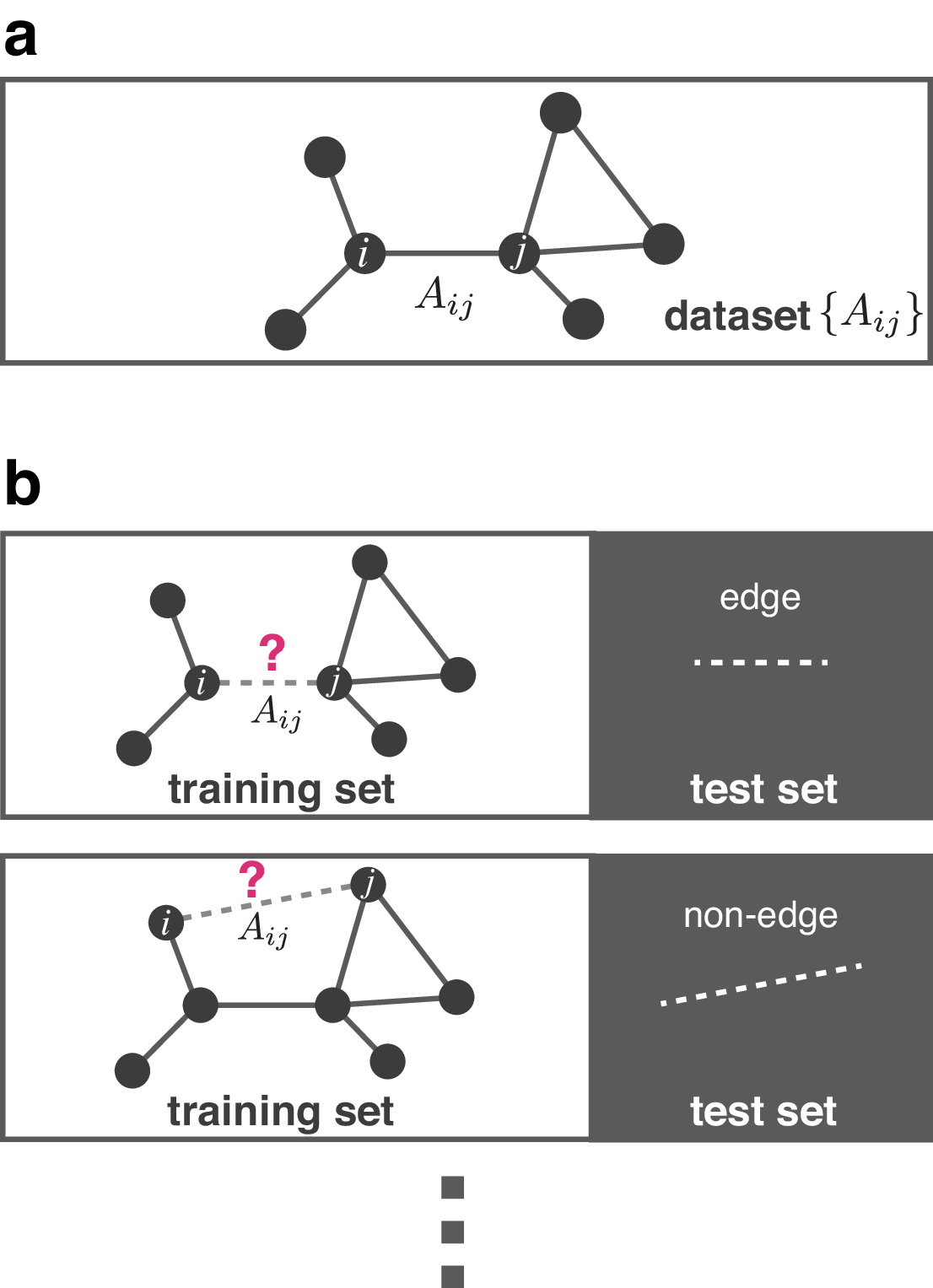}
 \end{center}
 \caption{\textbf{The LOOCV error estimate of the network data.} (a) The original dataset is given as the adjacency matrix $\ket{A}$ of a network. (b) In each prediction, we hide a single piece of edge or non-edge information from $\ket{A}$: this unobserved edge or unobserved non-edge is the test set, and the adjacency matrix $A^{\cav{i}{j}}$ without the information on that edge or non-edge is the training set. }
 \label{graphLOOCV}
\end{figure}

Generally speaking, the prediction error can be used as a measure for quantifying the generalisation ability of the model for unobserved data. 
In a data-rich situation, we can split the dataset into two parts, the training and test sets, where the training set is used to learn the model parameters and the test set is used to measure the model's prediction accuracy. 
We conduct this process for models that have different complexities, and we select the model that has the least prediction error or the model that is the most parsimonious if the least prediction error is shared by multiple models. 
In practice, however, the dataset is often insufficient to be used to conduct this process reliably. 
The cross-validation estimate is a way to overcome such a data-scarce situation. 
In $K$-fold cross-validation, for example, we randomly split the dataset into $K$ $(>1)$ subsets, keep one of them as a test set while the remaining subsets are used as the training sets, and measure the prediction error. 
Then, we switch the role of the test set to the training set, pick one of the training sets as a new test set, and measure the prediction error again. 
We repeat this process until every subset is used as a test set; then, we repeat the whole process for different random $K$-splits. 
The average over all the cases is the final estimate of the prediction error of a given model's complexity. 

For a dataset with $N$ elements, the $N$-fold cross-validation is called the \textit{leave-one-out cross-validation} (LOOCV); hereafter, we focus on this approach. 
Note that the method of splitting is unique in the LOOCV. 
In the context of a network, the dataset to be split is the set of edges and non-edges, and the data-splitting procedure is illustrated in Fig.~\ref{graphLOOCV}: for each pair of vertices, we learn the parameters while using the data without the information of edge existence between the vertex pair; then, we measure whether we can correctly predict the existence or non-existence of an edge. 
At first glance, this approach appears to be inefficient and redundant because we perform the parameter learning $N(N-1)/2$ times for similar training sets, which is true when the LOOCV is implemented in a straightforward manner. 
Nevertheless, we show that such a redundant process is not necessary and that the prediction error can be estimated efficiently using a BP-based algorithm. 
It is possible to extend the LOOCV estimate that we propose to the $K$-fold cross-validation. As we mention in the Methods section, however, such an estimate has both conceptual and computational issues, i.e., the LOOCV is an exceptionally convenient case.

The BP-based algorithm that we use to infer the cluster assignment was introduced by Decelle et al. \cite{Decelle2011,Decelle2011a}. 
This algorithm is scalable and performs well for sparse networks; this is favourable, because real-world networks are typically sparse. 
Indeed, it is known that the BP-algorithm asymptotically achieves the information-theoretic bound \cite{Mossel2014,Massoulie2014} (when the number of clusters is 2) in the sense of accuracy when the network is actually generated from the assumed model, the stochastic block model. 
In the original algorithm, the number of clusters $q^{\ast}$ is determined by the Bethe free energy, which is equivalent to the approximate negative marginalised log-likelihood; among the models of different (maximum) numbers of clusters, the most parsimonious model with low Bethe free energy is selected. 
Despite its plausibility, however, it is known that this prescription performs poorly in practice. 
One of the reasons is that the estimate using the Bethe free energy relies excessively on the assumption that the network is generated from the stochastic block model, which is almost always not precisely correct. 
We show that this issue can be substantially improved by evaluating cross-validation errors instead of the Bethe free energy while keeping the other parts of the algorithm the same. 
With this improvement, we can conclude that the BP-based algorithm is not only an excellent tool in the ideal case but is also useful in practice. 

%We denote the number of vertices in the networks as $N$. 
We denote the set of vertices and edges in the networks as $V$ and $E$, and their cardinalities as $N$ and $L$, respectively. 
Throughout the current study, we consider undirected sparse networks, and we ignore multi-edges and self-loops for simplicity. 
%We denote by $V$ and $E$ the sets of vertices and edges in the network, respectively, and refer to their corresponding cardinalities as $N$ and $L$. 

\section*{Results}\label{Results}

\subsection*{Stochastic block model}\label{StochasticBlockModel}
Let us first explain how an instance of the stochastic block model is generated. 
The stochastic block model is a random graph model that has a planted modular structure. 
The fraction of vertices that belong to cluster $\sigma$ is specified by $\gamma_{\sigma}$; accordingly, each vertex $i$ has its planted cluster assignment $\sigma_{i} \in \{1, \dots, q^{\ast}\}$, where $q^{\ast}$ is the number of clusters. 
The set of edges $E$ is generated by connecting pairs of vertices independently and randomly according to their cluster assignments. 
In other words, vertices $i$ and $j$ with cluster assignments $\sigma_{i} = \sigma$ and $\sigma_{j} = \sigma^{\prime}$ are connected with probability $\omega_{\sigma\sigma^{\prime}}$; the matrix $\ket{\omega}$ that specifies the connection probabilities within and between the clusters is called the affinity matrix. 
Note that the affinity matrix $\ket{\omega}$ is of $O(N^{-1})$ when the network is sparse. 

As the output, we obtain the adjacency matrix $\ket{A}$. 
While the generation of network instances is a forward problem, graph clustering is its inverse problem. 
In other words, we infer the cluster assignments of vertices $\ket{\sigma}$ or their probability distributions as we learn the parameters $(\ket{\gamma}, \ket{\omega})$, given the adjacency matrix $\ket{A}$ and the input (maximum) number of clusters $q$. 
After we have obtained the results for various values of $q$, we select $q^{\ast}$. 
Note here that the input value $q$ is the maximum number of clusters allowed and that the actual number of clusters that appears for a given $q$ can be less than $q$.

\subsection*{Cross-validation errors}
We consider four types of cross-validation errors. 
All of the errors are calculated using the results of inferences based on the stochastic block model. 
We denote $A^{\cav{i}{j}}$ as the adjacency matrix of a network in which $A_{ij}$ is unobserved, i.e., in which it is unknown whether $A_{ij}=0$ or $A_{ij}=1$. 
We hereafter generally denote $p(X \lvert Y)$ as the probability of $X$ that is conditioned on $Y$ or the likelihood of $Y$ when $X$ is observed. 

The process of edge prediction is two-fold: first, we estimate the cluster assignments of a vertex pair; then, we predict whether an edge exists. 
Thus, the posterior predictive distribution $p(A_{ij} = 1 | A^{\cav{i}{j}})$ of the model in which the vertices $i$ and $j$ are connected given dataset $A^{\cav{i}{j}}$, or the marginal likelihood of the learned model for the vertex pair, is the following: 
\begin{align}\label{MarginalEdgePrediction}
p(A_{ij} = 1 \lvert A^{\cav{i}{j}}) 
&= \sum_{\sigma_{i}, \sigma_{j}} p(A_{ij} = 1 \lvert \sigma_{i}, \sigma_{j}) p(\sigma_{i}, \sigma_{j} \lvert A^{\cav{i}{j}}) \notag\\
&= \bracket{p(A_{ij} = 1 \lvert \sigma_{i}, \sigma_{j})}_{A^{\cav{i}{j}}}, 
\end{align}
where $\bracket{\cdots}_{A^{\cav{i}{j}}}$ is the average over $p(\sigma_{i}, \sigma_{j} \lvert A^{\cav{i}{j}})$ and we have omitted some of the conditioned parameters. 
The error should be small when the prediction of edge existence for every vertex pair is accurate. 
In other words, the probability distribution in which each element has probability of equation (\ref{MarginalEdgePrediction}) is close, in the sense of the Kullback--Leibler (KL) divergence, to the actual distribution, which is given as the empirical distribution. 
Therefore, it is natural to employ, as a measure of the prediction error, the cross-entropy error function \cite{BishopPRML} 
\begin{align}
E_{\mathrm{Bayes}}(q) &= -\overline{\log p(A_{ij} \lvert A^{\cav{i}{j}})} \notag\\
%-\frac{1}{N} \sum_{i<j} \log p(A_{ij} \lvert p(\hat{A}_{ij} | A^{\cav{i}{j}}) ) \\
&= -\frac{1}{L} \sum_{i<j} \biggl[ A_{ij} \log p(A_{ij} = 1 \lvert A^{\cav{i}{j}}) \notag\\
&\hspace{10pt}+ (1-A_{ij}) \log p(A_{ij} = 0 \lvert A^{\cav{i}{j}}) \biggr], \label{EBayes}
\end{align}
where $\overline{X} \equiv \sum_{i<j} X(A_{ij})/L$. 
Note that we have chosen the normalisation in such a way that $E_{\mathrm{Bayes}}$ is typically $O(1)$ in sparse networks. 
We refer to equation (\ref{EBayes}) as the Bayes prediction error, which corresponds to the LOOCV estimate of the \textit{stochastic complexity} \cite{Levin1990}. 
As long as the model that we use is consistent with the data, the posterior predictive distribution is optimal as an element of the prediction error because the intermediate dependence $(\sigma_{i},\sigma_{j})$ is fully marginalised and gives the smallest error. 

Unfortunately, the assumption that the model that we use is consistent with the data is often invalid in practice. In that case, the Bayes prediction error $E_{\mathrm{Bayes}}$ may no longer be optimal for prediction. 
In equation (\ref{EBayes}), we employed $-\log p(A_{ij} \lvert A^{\cav{i}{j}})$ as the error of a vertex pair. 
Instead, we can consider the log-likelihood of cluster assignments $-\log p(A_{ij} \lvert \sigma_{i}, \sigma_{j})$ to be a fundamental element and measure $\bracket{-\log p(A_{ij} \lvert \sigma_{i}, \sigma_{j})}_{A^{\cav{i}{j}}}$ as the prediction error of a vertex pair.  In other words, the cluster assignments $(\sigma_{i}, \sigma_{j})$ are drawn from the posterior distribution, and the error of the vertex pair is measured with respect to those fixed assignments. 
Then, the corresponding cross-entropy error function is
\begin{align}
E_{\mathrm{Gibbs}}(q)
&= \overline{\bracket{-\log p(A_{ij} \lvert \sigma_{i}, \sigma_{j})}_{A^{\cav{i}{j}}}}. \label{EGibbs}
\end{align}
We refer to equation (\ref{EGibbs}) as the Gibbs prediction error. 
If the probability distribution of the cluster assignments is highly peaked, then $E_{\mathrm{Gibbs}}$ will be close to $E_{\mathrm{Bayes}}$, and both will be relatively small if those assignments predict the actual network well. % IF THE PREDICTION IS NOT SUFFICIENTLY ACCURATE, THEN THE ERROR WILL NOT BE SMALL EVEN IF THE DISTRIBUTION IS STRONGLY PEAKED. 
Alternatively, we can measure the \textit{maximum a posteriori} (MAP) estimate of equation (\ref{EGibbs}). 
In other words, instead of taking the average over $p(\sigma_{i}, \sigma_{j} \lvert A^{\cav{i}{j}})$, we select the most likely assignments to measure the error. 
We refer to $E_{\mathrm{MAP}}(q)$ as the prediction error. 

Finally, we define the Gibbs training error, which we refer to as $E_{\mathrm{training}}$. 
This error can be obtained by taking the average over $p(\sigma_{i}, \sigma_{j} \lvert A)$ instead in equation (\ref{EGibbs}), i.e., 
\begin{align}
E_{\mathrm{training}}(q)
&= \overline{\bracket{-\log p(A_{ij} \lvert \sigma_{i}, \sigma_{j})}_{A}}. \label{Etraining}
\end{align}
Because we make use of the information regarding the edge existence when we take the average, the result is not a prediction error, but is the goodness of fit of the assumed model to the given data. 

At first glance, the cross-validation errors that we presented above might appear computationally infeasible because we must know $p(\sigma_{i}, \sigma_{j} \lvert A^{\cav{i}{j}})$ with respect to every vertex pair. 
In the Methods section, however, we show that we have analytical expressions of the cross-validation errors in terms of the outputs of BP; therefore, the model assessment for the sparse networks is very efficient. 

%One might doubt that the consistency of the model with the actual data is still required in the present framework because we use the stochastic block model to assess the errors. Note that the goal here is not to identify the correct parameters of the true generative model, but to choose a model with high predictability among the stochastic block models of different complexity.

In the following subsections, we show the performances of these cross-validation errors for various networks relative to the performance of the Bethe free energy.

\subsection*{Real-world networks} 
\begin{figure*}[t]
 \begin{center}
    \includegraphics[width=0.99 \columnwidth]{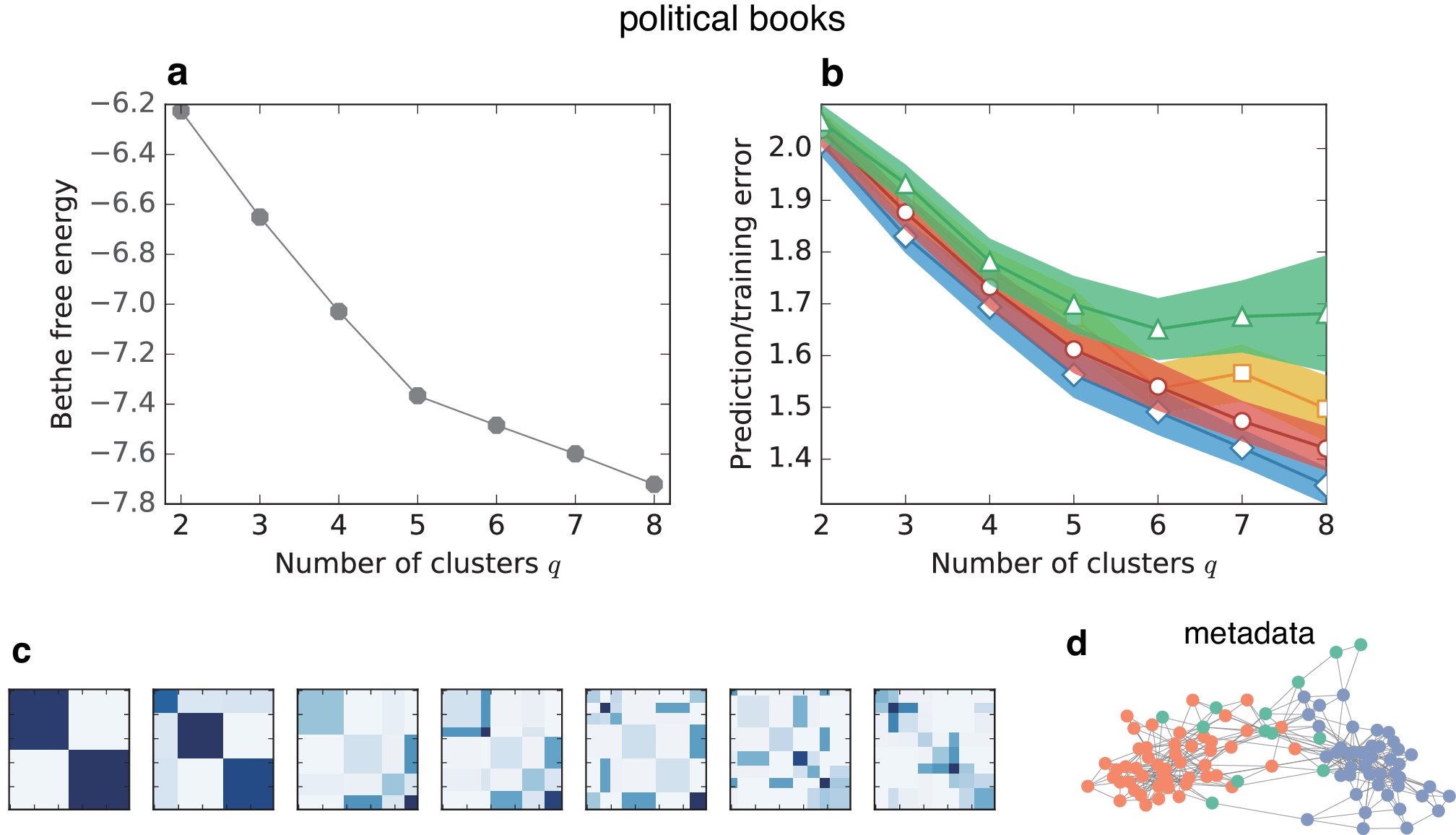}
 \end{center}
 \caption{
 	\textbf{Model assessments and inferred clusters in the network of books about US politics.} 
	(a) Bethe free energy and (b) prediction and training errors as functions of the number of clusters $q$. 
	The four data in (b) are, from top to bottom, 
	Gibbs prediction errors $E_{\mathrm{Gibbs}}$ (green triangles), 
	MAP estimates $E_{\mathrm{MAP}}$ of $E_{\mathrm{Gibbs}}$ (yellow squares), 
	Bayes prediction errors $E_{\mathrm{Bayes}}$ (red circles), and 
	Gibbs training errors $E_{\mathrm{training}}$ (blue diamonds). 
	For each error, the constant term is neglected and the standard error is shown in shadow. 
	(c) The learned parameters, $\ket{\omega}$ and $\ket{\gamma}$, are visualised from $q=2$ to $8$. %: each square represents the affinity matrix, where the size of each element reflects the learned cluster size, and the darker colour represents higher values of the affinity matrix element. 
	(d) The cluster assignments that are indicated in the metadata of the dataset; the vertices with the same colour belong to the same cluster. 
}
 \label{fig-politicalbooks}
\end{figure*}

\begin{figure*}[t]
 \begin{center}
   \includegraphics[width=0.95 \columnwidth]{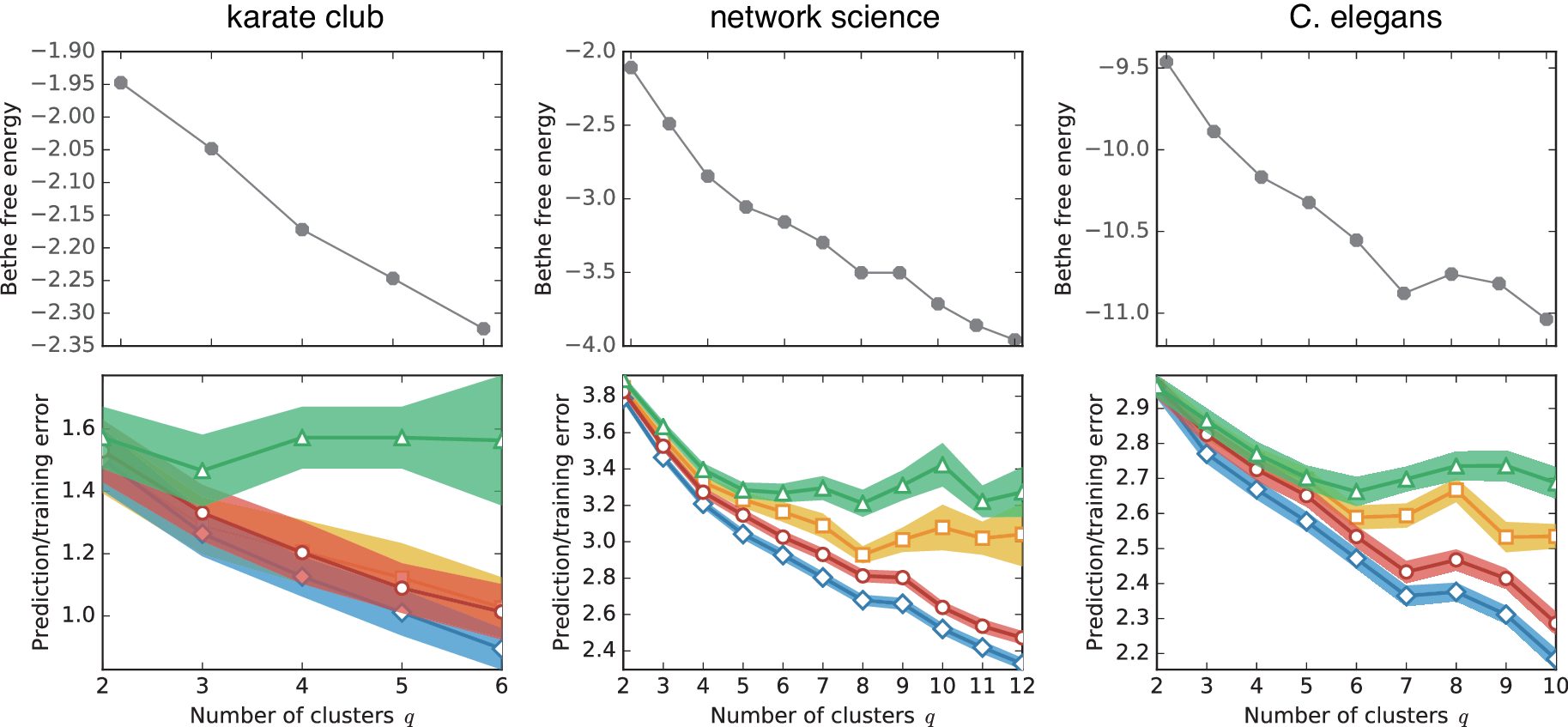}
 \end{center}
 \caption{
 \textbf{Results of model assessments for various real-world networks.} 
 They are plotted in the same manner as Figs.~\ref{fig-politicalbooks}a and \ref{fig-politicalbooks}b.
 }
 \label{fig-otherrealnetworks}
\end{figure*}

First, and most importantly, we show the performance of the Bethe free energy and cross-validation errors on real-world networks. 
Figures \ref{fig-politicalbooks-complexity} and \ref{fig-politicalbooks} show the results on the network of books on US politics \cite{Newman2006politicalbooks} (which we refer to as \textit{political books}). 
This network is a copurchase network whose vertices are books sold on Amazon.com, and each edge represents a pair of books that was purchased by the same buyer; 
the metadata of the dataset has the labels ``conservative'', ``liberal'', and ``neutral.'' 

Figure \ref{fig-politicalbooks}b shows that the cross-validation estimate of the Gibbs prediction error $E_{\mathrm{Gibbs}}$ saturates favourably, while the Bethe free energy (Fig.~\ref{fig-politicalbooks}a), Bayes prediction error $E_{\mathrm{Bayes}}$, and Gibbs training error $E_{\mathrm{training}}$ keep decreasing as we increase the input number of clusters $q$. 
Although the MAP estimate of the Gibbs prediction error $E_{\mathrm{MAP}}$ often exhibits similar behaviour to the Gibbs prediction error $E_{\mathrm{Gibbs}}$, we observe in the results of other datasets that it does not appear to be distinctively superior to $E_{\mathrm{Gibbs}}$. 

%It makes sense that the Bayes prediction error $E_{\mathrm{Bayes}}$ behaves similarly to the Bethe free energy. 
Compared with the Gibbs prediction error $E_{\mathrm{Gibbs}}$, the Bayes prediction error $E_{\mathrm{Bayes}}$ appears to be more sensitive to the assumption that a network is generated by the stochastic block model, so that we rarely observe the clear saturation of the LOOCV error. 
In a subsection below, we show how overfitting and underfitting occur by deriving analytical expressions for the differences among the cross-validation errors. 

How should we select the number of clusters $q^{\ast}$ from the obtained plots of cross-validation errors? 
Although this problem is well defined when we select the best model, i.e., the model that has the least error, it is common to select the most parsimonious model instead of the best model in practice. 
Then, how do we determine the ``most parsimonious model'' from the results? 
This problem is not well defined, and there is no principled prescription obtained by consensus. 
%This problem is also related to the level of model complexity that a practitioner can afford. 

However, there is an empirical rule called the ``one-standard error'' rule \cite{HastieTibshiraniFriedman} that has often been used. 
Recall that each cross-validation estimate is given as an average error per edge; thus, we can also measure its standard error. 
The ``one-standard error'' rule suggests selecting the simplest model whose estimate is no more than one standard error above the estimate of the best model. 
In the case of the \textit{political books} network, the best model is the model in which $q=6$. Because the simplest model within the range of one standard error is $q=5$, it is our final choice.

The actual partition for each value of $q$ is shown in Figs.~\ref{fig-politicalbooks-complexity} and \ref{fig-politicalbooks}c. 
The cluster assignments indicated in the metadata is presented in Fig.~\ref{fig-politicalbooks}d for reference. 
In Fig.~\ref{fig-politicalbooks}c, the learned values of the parameters $\ket{\omega}$ and $\ket{\gamma}$ are visualised: a higher value element of the affinity matrix $\ket{\omega}$ is indicated in a darker colour, and the size of an element reflects the fraction of the cluster size $\gamma_{\sigma}$. 
For $q = 3$, we can identify two communities and a cluster that is connected evenly to those communities, as presumed in the metadata. 
For $q=5$, in addition, the sets of core vertices in each community are also detected. 
Note that recovering the labels in the metadata is not our goal. 
The metadata are not determined based on the network structure and are not the ground-truth partition \cite{Peel2016}. 

It should also be noted that we minimise the Bethe free energy in the cluster inference step in every case. 
When we select the number of clusters $q^{\ast}$, i.e., in the model selection step, we propose to use the cross-validation errors instead of the Bethe free energy. 

The results of other networks are shown in Fig.~\ref{fig-otherrealnetworks}. 
They are the friendship network of a karate club \cite{karateclub}, coauthorship network of scientists who work on networks \cite{adjnounnetworkscience} (see \cite{NewmanDataset} for details of the datasets), and the metabolic network of C. elegans \cite{celegans}. 
We refer to those as \textit{karate club}, \textit{network science}, and \textit{C. elegans}, respectively. 
The results are qualitatively similar to the \textit{political books} network. 
Note that the initial state dependency can be sensitive when the input number of clusters $q$ is large. 
Therefore, the results can be unstable in such a region. 

From the results in Figs.~\ref{fig-politicalbooks} and \ref{fig-otherrealnetworks}, one might conclude that the Gibbs prediction error $E_{\mathrm{Gibbs}}$ is the only useful criterion in practice. However, for example, when the holdout method is used instead of the LOOCV (see the Methods section), it can be confirmed that the Bayes prediction error $E_{\mathrm{Bayes}}$ also behaves reasonably.

\subsection*{The error for $q=1$}
For the \textit{karate club} network in Fig.~\ref{fig-otherrealnetworks}, we see that the errors keep decreasing except for the Gibbs prediction error $E_{\mathrm{Gibbs}}$. Recall that, for the stochastic block model with large $q$ and low average degree to be detectable, the network is required to exhibit sufficiently strong modular structure \cite{Decelle2011a}. 
Thus, although we do not a priori know which prediction error should be referred to, because the network has average degree smaller than 5, it is hard to believe that the errors other than $E_{\mathrm{Gibbs}}$ are appropriate. The Gibbs prediction error $E_{\mathrm{Gibbs}}$ has the smallest error at $q=3$, and the model in which $q=2$ has an error within the range of one standard error of $q=3$. 
However, one might suspect that the most parsimonious model is the model with $q=1$, i.e., the model that we should assume is a uniform random graph, and there is no statistically significant modular structure. 
In this case, the connection probability for an arbitrary vertex pair is determined by a single parameter, i.e., $p(A_{ij} \lvert \sigma_{i}\sigma_{j}) = \omega$, and it is simply $\omega = 2L/N(N-1)$. 
Moreover, there is no difference among the errors that we listed above. 
Hence, 
\begin{align}
E_{\mathrm{Bayes}} = E_{\mathrm{Gibbs}} = E_{\mathrm{MAP}} = E_{\mathrm{training}} 
\approx 1 - \log \omega. \label{q1errorSBM}
\end{align}
Here, we used the fact that $\omega = O(N^{-1})$ because we consider sparse networks. 
In every plot of cross-validation errors, the constant and $O(N^{-1})$ terms are neglected. 
The number of vertices and edges in the \textit{karate club} network are $N = 34$ and $L = 78$, respectively, i.e., $-\log \omega \simeq 1.97$, which is much larger than the errors for $q=2$; we thus conclude that the number of clusters $q^{\ast} = 2$.

\subsection*{Degree-corrected stochastic block model}

\begin{figure}[t!]
 \begin{center}
   \includegraphics[width=0.6 \columnwidth]{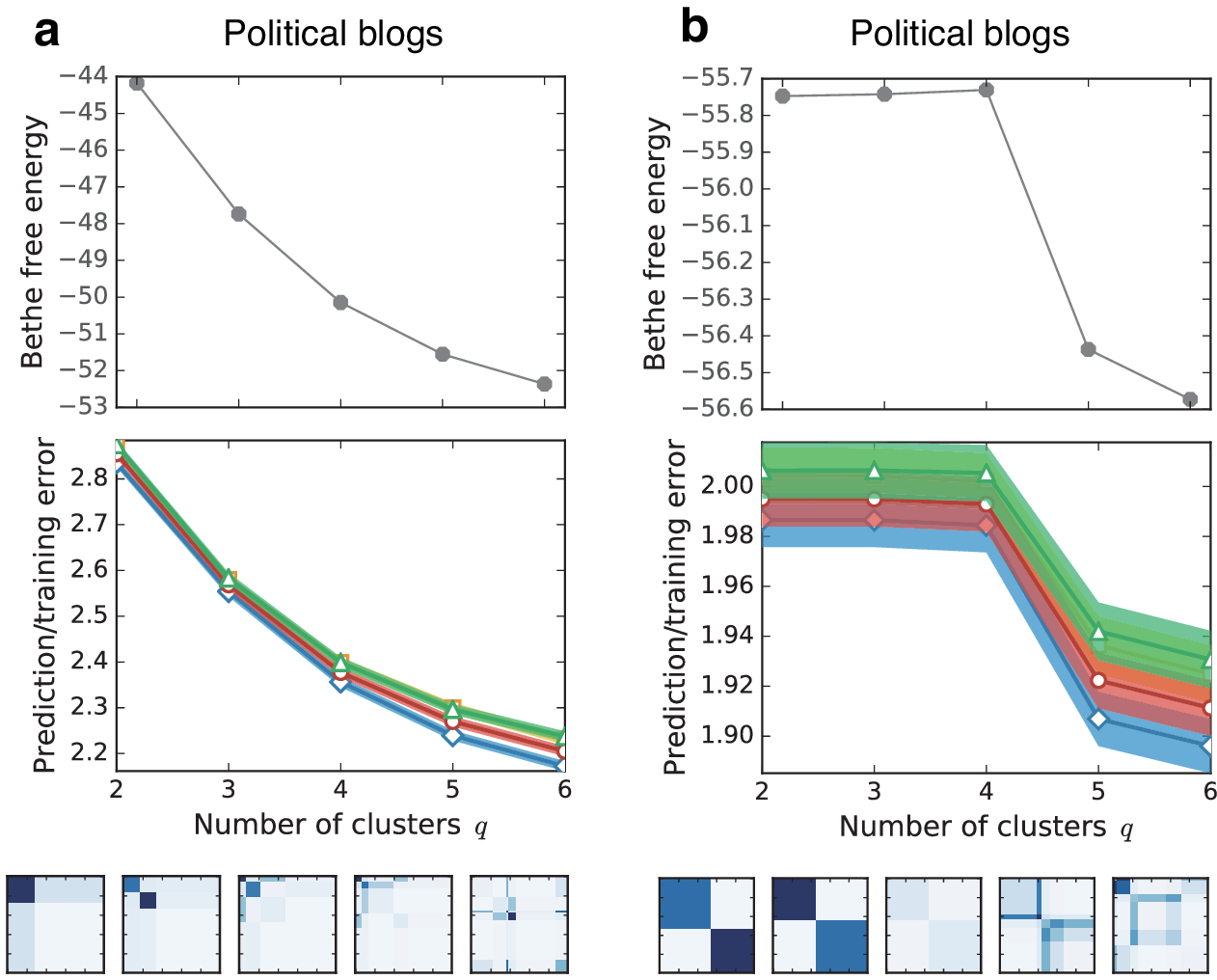}
 \end{center}
 \caption{
 \textbf{Results of the model assessments for the network of hyperlinks between blogs on US politics.} 
 (a) The results of the standard model and the learned parameters. 
 (b) The results of the degree-corrected model and the learned parameters. 
 }
 \label{dcSBMpoliticalblogs}
\end{figure}

It has been noted that the standard stochastic block model that we explained above is often not suitable for inferring clusters in real-world networks because many of them have a scale-free degree distribution, while the standard stochastic block model is restricted to having a Poisson degree distribution. 
This inconsistency affects both the cluster inference for a given $q^{\ast}$ and the model selection. 
For example, as shown in Fig.~\ref{dcSBMpoliticalblogs}a, we found that all of the criteria largely overfit for the \textit{political blogs} network when the standard stochastic block model is used. 
The \textit{political blogs} network is a network of hyperlinks between blogs on US politics (we neglect the directions of the edges), which is expected to have two clusters. 

The degree-corrected stochastic block model \cite{Karrer2011,ZhaoLevinaZhu2012} is often used as an alternative. 
This model has the degree-correction parameter $\theta_{i}$ for each vertex in addition to $\ket{\gamma}$ and $\ket{\omega}$, where $\ket{\theta}$ allows a cluster to have a heterogeneous degree distribution. (See \cite{Karrer2011} for details of the model.) 
The cross-validation model assessment can be straightforwardly extended to the degree-corrected stochastic block model: 
for the inference of cluster assignments, the BP algorithm can be found in \cite{Yan2014}; for the cross-validation errors, we only need to replace $\omega_{\sigma_{i}, \sigma_{j}}$ for the probability $p(A_{ij}=1 \lvert \sigma_{i}, \sigma_{j})$ with $\theta_{i}\omega_{\sigma_{i}, \sigma_{j}}\theta_{j}$ in each case. 
As shown in Fig.~\ref{dcSBMpoliticalblogs}b, the assessment is reasonable when a degree-corrected stochastic block model is used. 
Although the error drops for $q \ge 5$, it is better to discard this part because the numbers of iterations until convergence become relatively large there; we regard it as the ``wrong solution'' that is mentioned in Ref.~\cite{NewmanClauset2016}. 

When $q=1$, we do not assume the Erd{\H{o}}s-R\'{e}nyi random graph, but instead assume the random graph that has the same expected degree sequence as the actual dataset \cite{Newman2006PRE}, which is the model often used as the null model in modularity. 
Thus, the probability that vertices $i$ and $j$ are connected is $d_{i}d_{j}/2L$. 
After some algebra, assuming that $d_{i}d_{j}/2L \ll 1$, the error for $q=1$ is approximately 
\begin{align}
1 - \frac{1}{L} \sum_{i=1}^{N} d_{i} \log d_{i} + \log(2L), \label{q1errorDCSBM}
\end{align} 
which is equal to equation (\ref{q1errorSBM}) when the network is regular. 
In the case of the \textit{political blogs} network, the cross-validation error for $q=1$ is approximately $3.42$. (It is $2.42$ in the plot because the constant term is neglected.) 
Thus, we obtain $q^{\ast} = 2$.

\subsection*{Synthetic networks and the detectability threshold}

\begin{figure*}[t!]
 \begin{center}
   \includegraphics[width= \columnwidth]{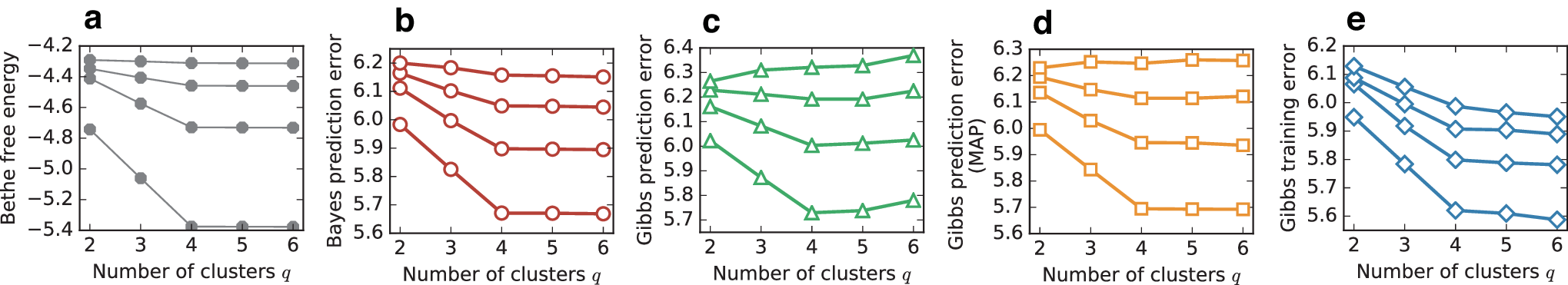}
 \end{center}
 \caption{
 	\textbf{Performance of the model assessment criteria on various stochastic block models.} 
	The results of (a) Bethe free energy, (b) Bayes prediction error $E_{\mathrm{Bayes}}$, (c) Gibbs prediction error $E_{\mathrm{Gibbs}}$, (d) MAP estimate of the Gibbs prediction error $E_{\mathrm{MAP}}$, and (e) Gibbs training error $E_{\mathrm{training}}$ are shown. 
	The lines in each plot represent the results of various values of $\epsilon$.
}
 \label{SBMs}
\end{figure*}

It is important to investigate how well the cross-validation errors perform in a critical situation, i.e., the case in which the network is close to the uniform random graph. 
To accomplish this goal, we observe the performance on the stochastic block model, the model that we assume for the inference itself. 
As explained in the subsection of the stochastic block model, the closeness to the uniform random graph is specified with the affinity matrix $\ket{\omega}$. 
Here, we consider a simple community structure: we set $\omega_{\mathrm{in}}$ for the diagonal elements, and the remaining elements are set to $\omega_{\mathrm{out}}$, where $\omega_{\mathrm{in}} \ge \omega_{\mathrm{out}}$. 
We parametrise the closeness to the uniform random graph, with $\epsilon = \omega_{\mathrm{out}}/\omega_{\mathrm{in}}$; $\epsilon = 0$ represents that the network consists of completely disconnected clusters, and $\epsilon = 1$ represents the uniform random graph. 
There is a phase transition point $\epsilon^{\ast}$ above which it is statistically impossible to infer a planted structure. This point is called the detectability threshold. 
Our interest here is to determine whether the planted number of clusters $q^{\ast}$ can be correctly identified using the cross-validation errors when $\epsilon$ is set to be close to $\epsilon^{\ast}$.

Figure \ref{SBMs} shows the Bethe free energy and cross-validation errors for the stochastic block models. 
In each plot, we set $q^{\ast} = 4$, each cluster has $1,000$ vertices, and the average degree $c$ is set to $8$. 
From bottom to top, the values of $\epsilon$ are $0.1$, $0.15$, $0.2$, and $0.25$, while the detectability threshold is $\epsilon^{\ast} \simeq 0.31$ \cite{Decelle2011a}. 
Note that when the value of $\epsilon$ is close to $\epsilon^{\ast}$, the inferred cluster assignments are barely correlated with the planted assignments; thus, the result is close to a random guess anyway. 

In all of the plots, the curves of validation become smoother as we increase the value of $\epsilon$, which indicates that it is difficult to identify the most parsimonious model. 
It is clear that the Bethe free energy and Bayes prediction error $E_{\mathrm{Bayes}}$ perform better than the Gibbs prediction error $E_{\mathrm{Gibbs}}$ in the present case because the networks we analyse correspond exactly to the model we assume. 
Figure \ref{SBMs} shows that the Gibbs prediction error $E_{\mathrm{Gibbs}}$ and its MAP estimate $E_{\mathrm{MAP}}$ underestimate $q^{\ast}$ near the detectability threshold. 
This finding is consistent with the results that we obtained for the real-world networks that the Bethe free energy and Bayes prediction error $E_{\mathrm{Bayes}}$ tend to overestimate. 
Indeed, under certain assumptions, we can derive that the Bayes prediction error $E_{\mathrm{Bayes}}$ identifies the planted number of clusters all the way down to the detectability threshold, while the Gibbs prediction error $E_{\mathrm{Gibbs}}$ strictly underfits near the detectability threshold. (See the Methods section for the derivation.)
Therefore, there is a trade-off between the Bayes prediction error $E_{\mathrm{Bayes}}$ and the Gibbs prediction error $E_{\mathrm{Gibbs}}$; their superiority depends on the accuracy of the assumption of the stochastic block model and the fuzziness of the network.

\subsection*{Relation among the cross-validation errors}
We show how the model assessment criteria that we consider in this study are related. 
First, we derive the relation among the errors $E_{\mathrm{Bayes}}$, $E_{\mathrm{Gibbs}}$, and $E_{\mathrm{training}}$. 
 By exploiting Bayes' rule, we have 
\begin{align}\label{ErrorRelation1}
\log p(A_{ij} \lvert A^{\cav{i}{j}}) &= \log p(A_{ij} \lvert \sigma_{i}, \sigma_{j}, A^{\cav{i}{j}}) \notag\\
&\hspace{30pt}+ \log \frac{p(\sigma_{i}, \sigma_{j} \lvert A^{\cav{i}{j}})}{p( \sigma_{i}, \sigma_{j} \lvert A)}. 
\end{align}
Note here that the left-hand side does not depend on $\sigma_{i}$ and $\sigma_{j}$. 
If we take the average with respect to $p(\sigma_{i}, \sigma_{j} \lvert A^{\cav{i}{j}})$ on both sides, 
\begin{align}\label{ErrorRelation2}
\log p(A_{ij} \lvert A^{\cav{i}{j}}) 
&= \bracket{ \log p(A_{ij} \lvert \sigma_{i}, \sigma_{j}, A^{\cav{i}{j}}) } \notag\\
&\hspace{10pt} + D_{\mathrm{KL}}\left( p(\sigma_{i}, \sigma_{j} \lvert A^{\cav{i}{j}}) \mid\mid p( \sigma_{i}, \sigma_{j} \lvert A) \right), 
\end{align}
where $D_{\mathrm{KL}}(p \lvert\lvert q)$ is the KL divergence. 
Taking the sample average of the edges, we obtain 
\begin{align}\label{EBayes-vs-EGibbs}
E_{\mathrm{Bayes}} &= E_{\mathrm{Gibbs}} - \overline{D_{\mathrm{KL}}\left( p(\sigma_{i}, \sigma_{j} \lvert A^{\cav{i}{j}}) \lvert\lvert p( \sigma_{i}, \sigma_{j} \lvert A) \right)}. 
\end{align}
If we take the average over $p(\sigma_{i}, \sigma_{j} \lvert A)$ in equation (\ref{ErrorRelation2}) instead, 
\begin{align}\label{EBayes-vs-Etraining}
E_{\mathrm{Bayes}} &= E_{\mathrm{training}} + \overline{D_{\mathrm{KL}}\left( p( \sigma_{i}, \sigma_{j} \lvert A) \lvert\lvert p(\sigma_{i}, \sigma_{j} \lvert A^{\cav{i}{j}}) \right)}. 
\end{align}
Because the KL divergence is non-negative, we have 
\begin{align}\label{ErrorRelation}
E_{\mathrm{training}} \le E_{\mathrm{Bayes}} \le E_{\mathrm{Gibbs}}. 
\end{align} 
%\cite{FootnoteErrorRelation}. 
This inequality follows directly from Bayes' rule and applies broadly. 
Note, however, that the amounts of the errors do not indicate the relationships among the numbers of clusters selected. 

Under a natural assumption, we can also derive the inequality for the number of clusters selected. 
Let $q$ be the trial number of clusters. 
If the cluster assignment distributions for different values of $q$ constitute a hierarchical structure, i.e., a result with a small $q$ can be regarded as the coarse graining of a result with a larger $q$, then the information monotonicity \cite{Csiszar2008,Amari2010} of the KL divergence ensures that 
\begin{align}
q^{\ast}_{\mathrm{training}} \ge q^{\ast}_{\mathrm{Bayes}} \ge q^{\ast}_{\mathrm{Gibbs}}. \label{q-inequality}
\end{align}
%In other words, the Gibbs prediction error $E_{\mathrm{Gibbs}}$ estimates a smaller number of clusters $q$ than the Bayes prediction error $E_{\mathrm{Bayes}}$ and the Gibbs training error $E_{\mathrm{training}}$. 
In other words, the Bayes prediction error $E_{\mathrm{Bayes}}$ can only overfit when the Gibbs prediction error $E_{\mathrm{Gibbs}}$ estimates $q^{\ast}$ correctly, while the Gibbs prediction error $E_{\mathrm{Gibbs}}$ can only underfit when the Bayes prediction error $E_{\mathrm{Bayes}}$ estimates $q^{\ast}$ correctly. 
In addition, when the model predicts the edge existence relatively accurately, $E_{\mathrm{MAP}}$ is biased in such a way that the error becomes small. Therefore, $E_{\mathrm{MAP}}$ tends to be smaller than $E_{\mathrm{Gibbs}}$. 
As observed for real-world networks, $E_{\mathrm{Gibbs}}$ typically performs well while $E_{\mathrm{Bayes}}$ often overfits in practice; equation (\ref{EBayes-vs-EGibbs}) implies that detailed information about the difference in the cluster assignment distribution is often not relevant and simply causes overfitting. 

In addition to the relationship among the cross-validation errors, it is also fruitful to seek relationships between the Bethe free energy and the prediction errors. 
We found that it is possible to express the Bethe free energy in terms of the prediction errors. 
(See the Methods section for the derivation.)

\section*{Discussion} 
To determine the number of clusters, we both proposed LOOCV estimates of prediction errors using BP and revealed some of the theoretical relationships among the errors. 
They are principled and scalable and, as far as we examined, perform well in practice. 
Unlike the BIC-like criteria, the prediction errors do not require the model consistency to be applicable. 
Moreover, although we only treated the standard and degree-corrected stochastic block models, the applicability of our LOOCV estimates is not limited to these models. 
With an appropriate choice of block models, we expect that the present framework can also be used in the vast majority of real-world networks, such as directed, weighted, and bipartite networks and networks that have positive and negative edges. 
This is in contrast to some criteria that are limited to specific modular structures. 

The code for the results of the current study can be found at \cite{graphBIXurl}. (See the description therein for more details of the algorithm.) 

The selection of the number of clusters $q^{\ast}$ can sometimes be subtle. 
For example, as we briefly mentioned above, when the inference algorithm depends sensitively on the initial condition, the results of the LOOCV can be unstable; all of the algorithms share this type of difficulty as long as the non-convex optimisation problem is considered. 
Sometimes the LOOCV curve can be bumpy in such a way that it is difficult to determine the most parsimonious model. 
For this problem, we note that there is much information other than the prediction errors that help us to determine the number of clusters $q^{\ast}$, and we should take them into account. 
For example, we can stop the assessment or discard the result when (i) the number of iterations until convergence becomes large \cite{NewmanClauset2016}, as we have observed in Fig.~\ref{dcSBMpoliticalblogs}b, (ii) the resulting partition does not exhibit a significant behaviour, e.g., no clear pattern in the affinity matrix, or (iii) the actual number of clusters does not increase as $q$ increases. 
All of this information is available in the output of our code. 

It is possible that the difficulty of selecting the number of clusters $q^{\ast}$ arises in the model itself. 
Fitting with the stochastic block models is flexible; thus, the algorithm can infer not only the assortative and disassortative structures but also more complex structures. 
However, the flexibility also indicates that slightly different models might be able to fit the data as good as the most parsimonious model. 
As a result, we can obtain a gradually decreasing LOOCV curve for a broad range of $q$. 
Thus, there should be a trade-off between the flexibility of the model and the difficulty of the model assessment.

%Despite the fact that model assessment based on predictability is a well-accepted principle, to the best of our knowledge, it has not been discussed in the literature for modular structures in the Bayesian framework. 
Despite the fact that model assessment based on prediction accuracy is a well-accepted principle, to the best of our knowledge, the method that is applicable to large-scale modular networks in the Bayesian framework is discussed in this study for the first time. 
For better accuracy and scalability, we believe that there is more to be done. 
One might investigate whether the LOOCV errors perform, in some sense, better than other BIC-like criteria. 
The relations and tendencies of the LOOCV errors compared to other model assessment criteria would certainly provide interesting future work even though it is unlikely that we can generally conclude the superiority of the criteria over the others \cite{Arlot2010}. 
In practice, if resource allocation allows, it is always better to evaluate multiple criteria \cite{Domingos2012}.

\section*{Methods}\label{Methods}
\subsection*{Inference of the cluster assignments}
Let us first explain the inference algorithm of the cluster assignment using the stochastic block model when the number of clusters $q$ is given. 
For this part, we follow the algorithm introduced in \cite{Decelle2011,Decelle2011a}: the expectation--maximisation (EM) algorithm using BP. 

An instance of the stochastic block model is generated based on selected parameters, $(\ket{\gamma}, \ket{\omega})$, as explained in the subsection of the stochastic block model. 
Because the cluster assignment of each vertex and the edge generation between each vertex pair are determined independently and randomly in the stochastic block model, its likelihood is given by 
\begin{align}
p(A, \ket{\sigma} \lvert \ket{\gamma}, \ket{\omega}, q) 
&= \prod_{i=1}^{N} \gamma_{\sigma_{i}} \prod_{i<j} \omega_{\sigma_{i}\sigma_{j}}^{A_{ij}} \left( 1-\omega_{\sigma_{i}\sigma_{j}}\right)^{1-A_{ij}}. \label{SBMlikelihood}
\end{align}

The goal here is to minimise the free energy defined by 
$f = -N^{-1}\log\sum_{\ket{\sigma}^{\prime}} p(A, \ket{\sigma}^{\prime} \lvert \ket{\gamma},\ket{\omega})$, which is equivalent to maximising the marginal log-likelihood. 
Using the identity $p(A, \ket{\sigma} \lvert \ket{\gamma},\ket{\omega}) = p(\ket{\sigma} \lvert A, \ket{\gamma},\ket{\omega}) p(A \lvert \ket{\gamma},\ket{\omega})$, the free energy $f$ can be expressed as  
\begin{align}
f 
%&= -\frac{1}{N}\sum_{\ket{\sigma}^{\prime}} p(A, \ket{\sigma}^{\prime} \lvert \ket{\gamma},\ket{\omega}) 
%= -\frac{1}{N}p(A \lvert \ket{\gamma},\ket{\omega}) 
= -\frac{1}{N}\log\frac{p(A, \ket{\sigma} \lvert \ket{\gamma},\ket{\omega})}{p(\ket{\sigma} \lvert A, \ket{\gamma},\ket{\omega})} 
\end{align}
for an arbitrary $\ket{\sigma}$. 
Taking the average over a probability distribution $q(\ket{\sigma})$ on both sides, we obtain the following variational expression. 
\begin{align}
f &= -\frac{1}{N} \sum_{\ket{\sigma}} q(\ket{\sigma}) \log \frac{p(A, \ket{\sigma} \lvert \ket{\gamma},\ket{\omega})}{q(\ket{\sigma})} 
\frac{q(\ket{\sigma})}{p(\ket{\sigma} \lvert A, \ket{\gamma},\ket{\omega})} \notag\\
&= \frac{1}{N}\Biggl[ -\sum_{\ket{\sigma}} q(\ket{\sigma}) \log p(A, \ket{\sigma} \lvert \ket{\gamma},\ket{\omega}) 
+ \sum_{\ket{\sigma}} q(\ket{\sigma}) \log q(\ket{\sigma}) \notag\\
&\hspace{40pt} - D_{\mathrm{KL}}\left(q(\ket{\sigma}) \lvert\lvert p(\ket{\sigma} \lvert A, \ket{\gamma},\ket{\omega})\right) 
\Biggr]. \label{VariationalExpression}
\end{align}
When $q(\ket{\sigma})$ is equal to the posterior distribution $p(\ket{\sigma} \lvert A,\ket{\gamma},\ket{\omega})$, the KL divergence disappears, and we should obtain the minimum. We can interpret the first and second terms as corresponding to the internal energy and negative entropy, respectively, and equation (\ref{VariationalExpression}) as the thermodynamic relation of the free energy. 
For the first term, by substituting equation (\ref{SBMlikelihood}), we have
\begin{align}
&-\frac{1}{N} \sum_{\ket{\sigma}} q(\ket{\sigma}) \log p(A, \ket{\sigma} \lvert \ket{\gamma},\ket{\omega}) \notag\\
&= -\frac{1}{N} \sum_{\ket{\sigma}} q(\ket{\sigma}) 
\Biggl[ \sum_{i} \log \gamma_{\sigma_{i}} + \sum_{i<j} \biggl( A_{ij} \log \omega_{\sigma_{i}\sigma_{j}} \notag\\
&\hspace{80pt}+ (1-A_{ij}) \log(1-\omega_{\sigma_{i}\sigma_{j}}) \biggr) \Biggr] \notag\\
&= -\frac{1}{N} \sum_{i}\sum_{\sigma} q^{i}_{\sigma} \log \gamma_{\sigma} 
- \frac{1}{N} \sum_{(i,j)\in E} \sum_{\sigma\sigma^{\prime}} q^{ij}_{\sigma\sigma^{\prime}} \log \omega_{\sigma\sigma^{\prime}} \notag\\
&\hspace{80pt}+ \frac{c}{2} + O(N^{-1}), \label{InternalEnergy}
\end{align}
where $q^{i}_{\sigma} = \bracket{\delta_{\sigma \sigma_{i}}}_{\ket{\sigma}}$ and $q^{ij}_{\sigma\sigma^{\prime}} = \bracket{\delta_{\sigma, \sigma_{i}} \delta_{\sigma^{\prime} \sigma_{j}}}_{\ket{\sigma}}$ are the marginal probabilities, where $\delta_{xy}$ is the Kronecker delta and $\bracket{\cdots}_{\ket{\sigma}}$ is the average over $q(\ket{\sigma})$. 
Therefore, the free energy minimisation in the variational expression requires two operations: the inference of the marginal probabilities of the cluster assignments, $q^{i}_{\sigma}$ and $q^{ij}_{\sigma\sigma^{\prime}}$, and the learning of the parameters, $(\ket{\gamma}, \ket{\omega})$. 
In the EM algorithm, they are performed iteratively. 

For the E-step of the EM algorithm, we infer the marginal probabilities of the cluster assignments using the current estimates of the parameters. 
Whereas the exact marginalisation is computationally expensive in general, BP yields an accurate estimate quite efficiently when the network is sparse, i.e., the case in which the network is locally tree-like. 
Here, we do not go over the complete derivation of the BP equations, which can be found in \cite{MezardMontanari2009,Decelle2011a}. 
We denote $\psi^{i}_{\sigma}$ as the BP estimate of $q^{i}_{\sigma}$, which is obtained by 
\begin{align}
\psi^{i}_{\sigma} &= \frac{1}{Z^{i}} \gamma_{\sigma} \mathrm{e}^{-h_{\sigma}} \prod_{k \in \partial i} \left( \sum_{\sigma_{k}} \psi^{k \to i}_{\sigma_{k}} \omega_{\sigma_{k}\sigma} \right), \label{SBMfullBP}
\end{align}
where $Z^{i}$ is the normalisation factor with respect to the cluster assignment $\sigma$, 
$h_{\sigma} = \sum_{k=1}^{N} \sum_{\sigma_{k}} \psi^{k}_{\sigma_{k}} \omega_{\sigma_{k}\sigma}$, which is due to the effect of non-edges $(i,k) \notin E$, 
and $\partial i$ indicates the set of neighbouring vertices of $i$. 
In equation (\ref{SBMfullBP}), we also have the so-called cavity bias $\psi^{i \to j}_{\sigma}$ for an edge $(i,j) \in E$; 
the cavity bias is the marginal probability of vertex $i$ without the marginalisation from vertex $j$. 
Analogously to equation (\ref{SBMfullBP}), the update equation for the cavity bias is obtained by 
\begin{align}
\psi^{i \to j}_{\sigma} &= \frac{1}{Z^{i \to j}} \gamma_{\sigma} \mathrm{e}^{-h_{\sigma}} \prod_{k \in \partial i \backslash j} \left( \sum_{\sigma_{k}} \psi^{k \to i}_{\sigma_{k}} \omega_{\sigma_{k}\sigma} \right), \label{SBMBP}
\end{align}
where $\partial i \backslash j$ indicates the set of neighbouring vertices of $i$ except for $j$, and $Z^{i \to j}$ is the normalisation factor. 
For the BP estimate $\psi^{ij}_{\sigma\sigma^{\prime}}$ of the two-point marginal $q^{ij}_{\sigma\sigma^{\prime}}$ for $(i,j) \in E$, we have 
\begin{align}
\psi^{ij}_{\sigma\sigma^{\prime}} = \frac{1}{Z^{ij}} \psi^{i \to j}_{\sigma} \omega_{\sigma\sigma^{\prime}} \psi^{j \to i}_{\sigma^{\prime}}, \label{TwoPointMarginal}
\end{align}
where $Z^{ij}$ is the normalisation factor with respect to the assignments $\sigma$ and $\sigma^{\prime}$. 

With these marginals in hand, in the M-step of the EM algorithm, we update the estimate of the parameters to $\hat{\ket{\gamma}}$ and $\hat{\omega}$ as 
\begin{align}
& \hat{\gamma}_{\sigma} = \frac{1}{N} \sum_{i=1}^{N} \psi^{i}_{\sigma}, \label{SBMgammaUpdate}\\
& \hat{\omega}_{\sigma\sigma^{\prime}} = \frac{1}{N^{2} \gamma_{\sigma} \gamma_{\sigma^{\prime}}} 
\sum_{(i,j) \in E} \frac{\omega_{\sigma\sigma^{\prime}} (\psi^{i \to j}_{\sigma} \psi^{j \to i}_{\sigma^{\prime}} + \psi^{j \to i}_{\sigma} \psi^{i \to j}_{\sigma^{\prime}}) }{Z^{ij}}. \label{SBMomegaUpdate}
\end{align}
Here, we have used the fact that the network is undirected. 
These update rules can be obtained by simply taking the derivatives with respect to the parameters in equation (\ref{InternalEnergy}) with the normalisation constraint $\sum_{\sigma} \gamma_{\sigma} = 1$. 

We recursively compute equation (\ref{SBMBP}) and the parameter learning, equations (\ref{SBMgammaUpdate}) and (\ref{SBMomegaUpdate}), until convergence; then, we obtain the full marginal using equation (\ref{SBMfullBP}), which yields the estimates of the cluster assignments of the vertices.

\subsection*{Bethe free energy and derivation of the cross-validation errors}
In the algorithm above, because we use BP to estimate the marginal probabilities, we are no longer minimising the free energy. 
Instead, as an approximation of the free energy, we minimise the Bethe free energy, which is expressed as 
\begin{align}
f_{\mathrm{Bethe}} &= -\frac{1}{N} \sum_{i} \log Z^{i} + \frac{1}{N} \sum_{(i,j)\in E} \log Z^{ij} - \frac{c}{2}, \label{BetheFreeEnergy}
\end{align}
where $Z^{i}$ and $Z^{ij}$ are the normalisation factors that appeared in equations (\ref{SBMfullBP}) and (\ref{TwoPointMarginal}). 
As mentioned in the Introduction section, to determine the number of clusters $q^{\ast}$ from the Bethe free energy $f_{\mathrm{Bethe}}$, we select the most parsimonious model among the models that have low $f_{\mathrm{Bethe}}$. 
This approach corresponds to taking the maximum likelihood estimation of the parameters. 
%In \cite{Decelle2011a}, $-c \log N/2$ is added to equation (\ref{BetheFreeEnergy}) to make the Bethe free energy extensive \cite{FootnoteConvention}; 
%in the numerical experiments, we follow their convention. 

We now explain the derivations of the cross-validation errors that we used in the Results section. 
The LOOCV estimate of the Bayes prediction error $E_{\mathrm{Bayes}}$ is measured by equation (\ref{EBayes}), and its element $p(A_{ij} = 1 \lvert A^{\cav{i}{j}})$ is given as equation (\ref{MarginalEdgePrediction}). 
The first factor in the sum of equation (\ref{MarginalEdgePrediction}) is simply $p(A_{ij} = 1 \lvert \sigma_{i}, \sigma_{j}) = \omega_{\sigma_{i}\sigma_{j}}$, by the model definition. 
An important observation is that because the cavity bias $\psi^{i \to j}_{\sigma_{i}}$ represents the marginal probability of vertex $i$ without the information from vertex $j$, this entity is exactly what we need for prediction in the LOOCV, i.e., $p(\sigma_{i}, \sigma_{j} \lvert A^{\cav{i}{j}}) = \psi^{i \to j}_{\sigma_{i}} \psi^{j \to i}_{\sigma_{j}}$. 
Additionally, recall that $p(\sigma_{i}, \sigma_{j} \lvert A_{ij} = 1,A^{\cav{i}{j}}) = \psi^{i \to j}_{\sigma_{i}} \omega_{\sigma_{i}\sigma_{j}}\psi^{j \to i}_{\sigma_{j}}/Z^{ij}$, where $Z^{ij}$ is defined in equation (\ref{TwoPointMarginal}). 
Thus, we have $p(A_{ij} = 1 \lvert A^{\cav{i}{j}}) = Z^{ij}$. 
By using the fact that $L = O(N)$ and $p(A_{ij} = 1 \lvert A^{\cav{i}{j}}) = O(N^{-1})$, $E_{\mathrm{Bayes}}(q)$ can be written as 
\begin{align}
E_{\mathrm{Bayes}}(q) 
%&= -\frac{1}{L} \sum_{(i,j) \in E} \log p(A_{ij} = 1 \lvert A^{\cav{i}{j}}) + \mathrm{const.} + O(N^{-1}) \notag\\
%&= -\frac{1}{L} \sum_{(i,j) \in E} \log \sum_{\sigma_{i},\sigma_{j}} \psi^{i \to j}_{\sigma_{i}} \omega_{\sigma_{i}\sigma_{j}} \psi^{j \to i}_{\sigma_{j}} + \mathrm{const.} + O(N^{-1}) \notag\\
&= 1 - \frac{1}{L} \sum_{(i,j) \in E} \log Z^{ij} + O(N^{-1}). \label{EBayes2}
\end{align}
%Up to a multiplicative and an additive constant, this expression is the Bethe free energy, in which the contribution from the one-point partition functions $Z^{i}$ is neglected. (THE SIGN IS OPPOSITE; THUS, THEY ARE TOTALLY DIFFERENT.)
Equation (\ref{EBayes2}) indicates that the prediction with respect to the non-edges contributes only a constant; thus, $E_{\mathrm{Bayes}}(q)$ essentially measures whether the existence of the edges is correctly predicted in a sparse network. 

The Gibbs prediction error $E_{\mathrm{Gibbs}}$ in equation (\ref{EGibbs}) can be obtained in the same manner. 
Using the approximation that the network is sparse, it can be written in terms of the cavity biases as 
\begin{align}
E_{\mathrm{Gibbs}}(q)
&\simeq 1 - \frac{1}{L} \sum_{(i,j)\in E} \sum_{\sigma_{i}, \sigma_{j}} 
\psi^{i \to j}_{\sigma_{i}}  \psi^{j \to i}_{\sigma_{j}} \log \omega_{\sigma_{i} \sigma_{j}}. \label{EGibbs2} %\notag\\
%& \hspace{60pt} + \mathrm{const.} + O(N^{-1}). 
\end{align}
%Here, we have omitted the constant and the $O(N^{-1})$ terms. 
The (MAP) estimate of the Gibbs prediction error $E_{\mathrm{MAP}}(q)$ can be obtained by replacing $\psi^{i \to j}_{\sigma_{i}}$ with $\delta_{\sigma_{i}, \mathrm{argmax}\{\psi^{i \to j}_{\sigma}\}}$ in equation (\ref{EGibbs2}). 

For the Gibbs training error $E_{\mathrm{training}}$ in equation (\ref{Etraining}), we take the average over $p(\sigma_{i}, \sigma_{j} \lvert A)$ instead of $p(\sigma_{i}, \sigma_{j} \lvert A^{\cav{i}{j}})$. 
Thus, we have %omitting the constant and $O(N^{-1})$ terms,
\begin{align}
E_{\mathrm{training}}(q)
&\simeq 1 - \frac{1}{L} \sum_{(i,j)\in E} \sum_{\sigma_{i}, \sigma_{j}} 
\frac{\psi^{i \to j}_{\sigma_{i}} \omega_{\sigma_{i} \sigma_{j}} \psi^{j \to i}_{\sigma_{j}}}{Z^{ij}} \log \omega_{\sigma_{i} \sigma_{j}}. \label{Etraining2} %\notag\\
%&\hspace{60pt}+ \mathrm{const.} + O(N^{-1}),
\end{align}
This training error can be interpreted as a part of the internal energy because it corresponds to the BP estimate of the second term in equation (\ref{InternalEnergy}). 

Note that all of the cross-validation errors, equations (\ref{EBayes2})--(\ref{Etraining2}), are analytical expressions in terms of the parameters and cavity biases. 
Therefore, we can readily measure those errors by simply running the algorithm once. 
It should also be noted that using BP for the LOOCV itself is not totally new; this idea has been addressed in a different context in the literature, e.g., Ref.~\cite{Opper1996}.

\subsection*{Detectability of the Bayes and Gibbs prediction errors}
Let us evaluate the values of the Bayes and Gibbs prediction errors, $E_{\mathrm{Bayes}}$ and $E_{\mathrm{Gibbs}}$, for the stochastic block model with the simple community structure, i.e., $\omega_{\sigma \sigma^{\prime}} = (\omega_{\mathrm{in}} - \omega_{\mathrm{out}})\delta_{\sigma \sigma^{\prime}} + \omega_{\mathrm{out}}$. 
By substituting this affinity matrix to equations (\ref{EBayes2}) and (\ref{EGibbs2}), we obtain 
\begin{align}
E_{\mathrm{Bayes}}(q) &= 1 - \log \omega_{\mathrm{out}} \notag\\
& - \frac{1}{L} \sum_{(i,j) \in E} \log\left( 1 + \frac{\omega_{\mathrm{in}} - \omega_{\mathrm{out}}}{\omega_{\mathrm{out}}} \sum^{q}_{\sigma=1}\psi^{i \to j}_{\sigma}\psi^{j \to i}_{\sigma} \right)\label{Ebayes_q2}\\
E_{\mathrm{Gibbs}}(q) &= 1 - \log \omega_{\mathrm{out}} \notag\\
& -\left(\log\frac{\omega_{\mathrm{in}}}{\omega_{\mathrm{out}}} \right) \frac{1}{L} \sum_{(i,j) \in E} \sum^{q}_{\sigma=1} \psi^{i \to j}_{\sigma}\psi^{j \to i}_{\sigma}. \label{Egibbs_q2}
\end{align}

As we sweep the parameter $\epsilon = \omega_{\mathrm{out}}/\omega_{\mathrm{in}} \to 1$ in the stochastic block model that has 2 clusters, $q = 2$ is selected as long as the prediction error for $q=2$ is less than the error for $q=1$. 
When the number of vertices $N$ is sufficiently large, $\omega = \omega_{\mathrm{out}} + (\omega_{\mathrm{in}} - \omega_{\mathrm{out}}) \sum^{2}_{\sigma=1} \gamma^{2}_{\sigma}$ in equation (\ref{q1errorSBM}). 
Thus, 
\begin{align}
& E_{\mathrm{Bayes}}(q=1) = E_{\mathrm{Gibbs}}(q=1) \notag\\
&\hspace{20pt}= 1 - \log \omega_{\mathrm{out}} 
- \log\left( 1+\frac{\omega_{\mathrm{in}} - \omega_{\mathrm{out}}}{\omega_{\mathrm{out}}} \sum^{2}_{\sigma=1} \gamma^{2}_{\sigma} \right), 
\end{align}
where we neglected the $O(N^{-1})$ contribution. 
Let us presume that the cavity bias $\psi^{i \to j}_{\sigma}$ is correlated with the planted structure up to the detectability threshold $\epsilon^{\ast}$ and $\psi^{i \to j}_{\sigma} = \gamma_{\sigma}$ for $\epsilon \ge \epsilon^{\ast}$, which is rigorously proven \cite{Mossel2014,Massoulie2014} in the case of 2 clusters of equal size. 
We then have 
\begin{align}
E_{\mathrm{Bayes}}(q=2) = E_{\mathrm{Bayes}}(q=1) 
\end{align}
in the undetectable region. 
In the case of equal size clusters, $E_{\mathrm{Bayes}}(q=2)$ is strictly smaller thatn $E_{\mathrm{Bayes}}(q=1)$ in the detectable region because the last term in (\ref{Ebayes_q2}) is minimised when the distribution is uniform. 
In contrast, when $\psi^{i \to j}_{\sigma} = \gamma_{\sigma}$, the inequality $\log(1+xy)>y\log(1+x)$ ($x>0$, $0<y<1$) ensures that 
\begin{align}
E_{\mathrm{Gibbs}}(q=2) > E_{\mathrm{Gibbs}}(q=1).
\end{align}
Therefore, assuming that the parameters $\ket{\gamma}$ and $\ket{\omega}$ are learned accurately, the Bayes prediction error $E_{\mathrm{Bayes}}$ suggests $q^{\ast}=2$ all the way down to the detectability threshold, while the Gibbs prediction error $E_{\mathrm{Gibbs}}$ strictly underfits near the detectability threshold.

\subsection*{The Bethe free energy in terms of the prediction errors}
We can also express the Bethe free energy $f_{\mathrm{Bethe}}$ in terms of the prediction errors. 
Ignoring the constant terms and factors, we observe that the Bayes prediction error $E_{\mathrm{Bayes}}$, equation (\ref{EBayes2}), is one component of the Bethe free energy, equation (\ref{BetheFreeEnergy}). 
The remaining part is the contribution of $-\sum_{i} \log Z^{i}$ that arises as follows. 

Let us consider hiding a vertex and the edges and non-edges that are incident to that vertex, instead of an edge. 
We denote $\{A_{ij}\}_{j \in V}$ as the set of adjacency matrix elements that are incident to vertex $i$, and $A^{\backslash i}$ as the adjacency matrix in which the information of $\{A_{ij}\}_{j \in V}$ is unobserved. 
For the prediction of an edge's existence, for example, because the unobserved vertex receives no information about its cluster assignment from its neighbours, we have 
\begin{align}
p(A_{ij} = 1 \lvert A^{\backslash i}) &= \sum_{\sigma_{i}\sigma_{j}} p(A_{ij}=1 \lvert \sigma_{i}\sigma_{j}) p(\sigma_{i}\sigma_{j} \lvert A^{\backslash i}) \notag\\
&= \sum_{\sigma_{i}\sigma_{j}} \omega_{\sigma_{i}\sigma_{j}} \gamma_{\sigma_{i}} \psi^{j \to i}_{\sigma_{j}}. \label{LeaveNodeOut1}
\end{align}
When we consider the prediction of the set of edges $\{A_{ij}\}_{j \in V}$, we find that all of the vertex pairs share the same cluster assignment for vertex $i$. Therefore, using the approximation that the network is sparse, we have 
\begin{align}
p(\{A_{ij}\}_{j \in V} \lvert A^{\backslash i}) 
&= \sum_{\sigma_{i}} p(\sigma_{i} \lvert A^{\backslash i}) \notag\\
&\hspace{10pt}\times \prod_{j \in \partial i} \left( \sum_{\sigma_{j}} p(A_{ij}=1 \lvert \sigma_{i}\sigma_{j})p(\sigma_{j}\lvert A^{\backslash i}) \right) \notag\\
&\hspace{10pt}\times \prod_{k \notin \partial i} \left( \sum_{\sigma_{k}} p(A_{ik}=0 \lvert \sigma_{i}\sigma_{k})p(\sigma_{k}\lvert A^{\backslash i}) \right) \notag\\
%&= \sum_{\sigma_{i}} \gamma_{\sigma_{i}} \prod_{j \in \partial i} \left( \sum_{\sigma_{j}} \psi^{j \to i}_{\sigma_{j}} \omega_{\sigma_{j}\sigma_{i}} \right) \prod_{k \notin \partial i} \left( \sum_{\sigma_{k}} \psi^{k \to i}_{\sigma_{k}} \left( 1 - \omega_{\sigma_{k}\sigma_{i}} \right) \right) \\
&\approx \sum_{\sigma_{i}} \gamma_{\sigma_{i}} \mathrm{e}^{-h_{\sigma_{i}}} \prod_{j \in \partial i} \left( \sum_{\sigma_{j}} \psi^{j \to i}_{\sigma_{j}} \omega_{\sigma_{j}\sigma_{i}} \right) \notag\\
&= Z^{i}. \label{LeaveNodeOut2}
\end{align}
Again, due to the sparsity, the prediction for the non-edges happened to be neglected, and it is essentially the prediction of the edges. 
Analogously to equation (\ref{EBayes}), we can consider a \textit{leave-one-vertex-out} version of the Bayes prediction error, which we refer to as $E^{v}_{\mathrm{Bayes}}$, as follows. 
\begin{align}
&E^{v}_{\mathrm{Bayes}}(q) 
= -\frac{1}{L}\sum_{ij} \biggl[ A_{ij} \log p(A_{ij}=1 \lvert A^{\backslash i}) \notag\\ 
&\hspace{60pt} + (1-A_{ij}) \log p(A_{ij}=0 \lvert A^{\backslash i}) \biggr] \notag\\
&= -\frac{1}{L} \sum_{i} \log \left( \prod_{j \in \partial i} p(A_{ij}=1 \lvert A^{\backslash i}) \prod_{j \notin \partial i} p(A_{ij}=0 \lvert A^{\backslash i}) \right) \notag\\
&= -\frac{1}{L} \sum_{i} \log p(\{A_{ij}\}_{j \in V} \lvert A^{\backslash i}) \notag\\
&= -\frac{1}{L} \sum_{i} \log Z^{i}. \label{LeaveNodeOut3}
\end{align}
Again, the edges and non-edges that are incident to the unobserved vertex $i$ share the same cluster assignment $\sigma_{i}$ at one end. 
In this definition, however, the prediction of an edge's existence is made twice for every edge. 
If the Bayes prediction error $E_{\mathrm{Bayes}}$ is subtracted to prevent this over-counting, it is exactly the Bethe free energy $f_{\mathrm{Bethe}}$ up to the constant terms and the normalisation factor.

\subsection*{Derivation of inequality (\ref{q-inequality})}
\begin{figure}[t]
 \begin{center}
  \includegraphics[width=0.5 \columnwidth]{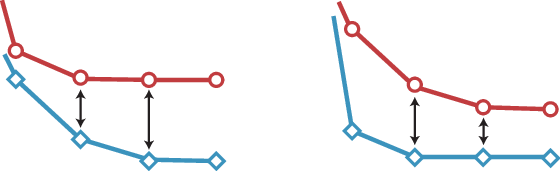}
 \end{center}
 \caption{\textbf{Schematic of the shapes of the error functions.} 
 As long as the gap between two error functions is nondecreasing, the error with the smaller value does not saturate earlier than the other. 
 }
 \label{qSelection}
\end{figure}

Recall that the Bayes prediction error $E_{\mathrm{Bayes}}$, Gibbs prediction error $E_{\mathrm{Gibbs}}$, and Gibbs training error $E_{\mathrm{training}}$ are related via equations (\ref{EBayes-vs-EGibbs}) and (\ref{EBayes-vs-Etraining}). 
We select the number of clusters $q$ as the point at which the error function saturates (i.e., stops decreasing) with increasing $q$. 
For a smaller $q$ to be selected by $E_{\mathrm{Bayes}}$ than by $E_{\mathrm{Gibbs}}$, the gap between them, $\overline{D_{\mathrm{KL}}\left( p(\sigma_{i}, \sigma_{j} \lvert A^{\cav{i}{j}}) \lvert\lvert p( \sigma_{i}, \sigma_{j} \lvert A) \right)}$, must decrease (see Fig.~\ref{qSelection}). 
In this subsection, we explain the information monotonicity of the KL divergence and when it is applicable in the present context. 

\begin{figure*}[t]
 \begin{center}
  \includegraphics[width=0.99 \columnwidth]{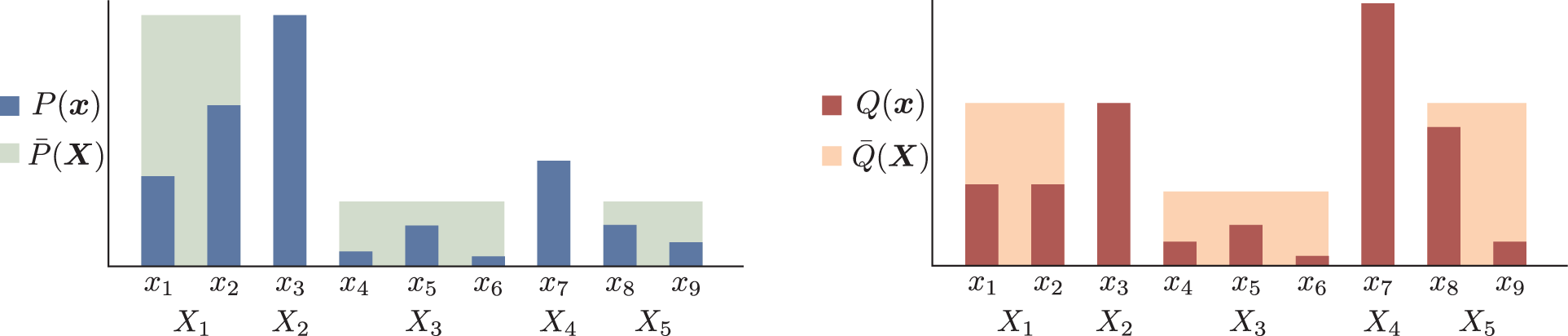}
 \end{center}
 \caption{
 \textbf{Example of the refinement of the probability distributions.} 
 We can regard $P(\ket{x})$ and $Q(\ket{x})$ as the refinements of $\bar{P}(\ket{X})$ and $\bar{Q}(\ket{X})$, respectively, with 
 $\ket{x}^{1} = \{ x_{1},x_{2} \}$, 
 $\ket{x}^{2} = \{ x_{3} \}$, 
 $\ket{x}^{3} = \{ x_{4},x_{5},x_{6} \}$, 
 $\ket{x}^{4} = \{ x_{7} \}$, and 
 $\ket{x}^{5} = \{ x_{8},x_{9} \}$ as a possible correspondence. 
 Note that the correspondence is not unique. 
 If we refer only to $Q(\ket{x})$ and $\bar{Q}(\ket{X})$, the assignment of $\{x_{1},x_{2}\}$, $\{x_{3}\}$, and $\{x_{8},x_{9}\}$ is exchangeable within $X_{1}$, $X_{2}$, and $X_{5}$. 
 However, $\bar{P}(X_{5})$ does not coincide with $P(\{ x_{1},x_{2} \})$ or $P(x_{3})$; therefore, only $X_{1}$ and $X_{2}$ are exchangeable between $\{x_{1},x_{2}\}$ and $\{x_{3}\}$. 
 The same argument applies to $\{x_{4},x_{5},x_{6}\}$ and $\{x_{8},x_{9}\}$. 
 }
 \label{ProbabilityRefinementFIg}
\end{figure*}

Let us consider sets of variables $\ket{X} = \{X_{1}, \dots, X_{m}\}$ and $\ket{x} = \{x_{1},\dots,x_{n}\}$ ($n > m$). 
We first set a pair of probability distributions $\bar{P}(\ket{X})$ and $\bar{Q}(\ket{X})$ on $\ket{X}$. 
Then we define a pair of probability distributions $P(\ket{x})$ and $Q(\ket{x})$ on $\ket{x}$ as refinements of $\bar{P}(\ket{X})$ and $\bar{Q}(\ket{X})$, respectively,
if there exists a family of sets $\{\ket{x}^{\mu}\}_{\mu=1}^{m}$ that is a partition of $\ket{x}$, i.e., $\ket{x}^{\mu} \cap \ket{x}^{\mu^{\prime}} = \emptyset$ for $\mu\ne\mu^{\prime}$ and $\cup_{\mu} \ket{x}^{\mu} = \ket{x}$, that satisfies $P(\ket{x}^{\mu}) = \bar{P}(X_{\mu})$ and $Q(\ket{x}^{\mu}) = \bar{Q}(X_{\mu})$ for any $\mu$. 
In other words, $\ket{X}$ can be regarded as the coarse graining of $\ket{x}$. 
An example is given in Fig.~\ref{ProbabilityRefinementFIg}. 
Note, however, that if $\ket{X}$ is actually constructed as the coarse graining of $\ket{x}$, the above condition trivially holds. %for $\bar{P}=P$ and $\bar{Q}=Q$. 
In general, a family that satisfies the above condition might not exist; even if it exists, it might not be unique.

The \textit{information monotonicity} of the KL divergence states that for $P(\ket{x})$ and $Q(\ket{x})$, which are the refinements of $\bar{P}(\ket{X})$ and $\bar{Q}(\ket{X})$, respectively, we have
\begin{align}\label{Monotonicity}
D_{\mathrm{KL}}(P \lvert\lvert Q) \ge D_{\mathrm{KL}}(\bar{P} \lvert\lvert \bar{Q}), 
\end{align}
which is natural, because the difference between the distributions is more visible at a finer resolution. 
Equation (\ref{Monotonicity}) is deduced by the convexity of the KL divergence. 
First, we can rewrite the right-hand side of equation (\ref{Monotonicity}) in terms of $P$ and $Q$ as follows: 
\begin{align}
D_{\mathrm{KL}}(\bar{P} \lvert\lvert \bar{Q}) = \sum_{\mu=1}^{m} P(\ket{x}^{\mu}) \log \frac{P(\ket{x}^{\mu})}{Q(\ket{x}^{\mu})}; 
\end{align}
Thus, if 
\begin{align}\label{elementMonotonicity}
\sum_{x\in x^{\mu}} P(x) \log \frac{P(x)}{Q(x)} \ge P(\ket{x}^{\mu}) \log \frac{P(\ket{x}^{\mu})}{Q(\ket{x}^{\mu})} 
\end{align}
holds for an arbitrary $\mu$, then equation (\ref{Monotonicity}) holds. 
We split $\ket{x}^{\mu}$ into $x_{1} \in \ket{x}^{\mu}$ and $\ket{x}^{\mu}\backslash x_{1}$, 
and we denote the corresponding probabilities as $P_{1} := P(x_{1})$, $Q_{1} := Q(x_{1})$, $P^{c}_{1} := P(\ket{x}^{\mu}\backslash x_{1})$, and $Q^{c}_{1} := Q(\ket{x}^{\mu}\backslash x_{1})$. 
The right-hand side of equation (\ref{elementMonotonicity}) is then 
\begin{align}\label{elementMonotonicity2}
%p(\ket{x}^{\mu}) \log \frac{p(\ket{x}^{\mu})}{q(\ket{x}^{\mu})} 
&\left( P_{1} + P^{c}_{1} \right) \log\left(\frac{P_{1} + P^{c}_{1}}{Q_{1} + Q^{c}_{1}}\right) \notag\\
&= -\left( P_{1} + P^{c}_{1} \right) \log\left(\frac{P_{1}}{P_{1} + P^{c}_{1}} \frac{Q_{1}}{P_{1}} + \frac{P^{c}_{1}}{P_{1} + P^{c}_{1}} \frac{Q^{c}_{1}}{P^{c}_{1}} \right) \notag\\
& \le P_{1} \log \frac{P_{1}}{Q_{1}} + P^{c}_{1} \log \frac{P^{c}_{1}}{Q^{c}_{1}}, 
\end{align}
where we used the convexity of the logarithmic function. 
By repeating the same argument for the second term of equation (\ref{elementMonotonicity2}), we obtain equation (\ref{elementMonotonicity}). 
Although the KL divergence is our focus, the information monotonicity holds more generally, e.g., for $f$-divergence \cite{Amari2010}.

We now use the information monotonicity to estimate the error functions. 
In the present context, the sets of variables $\ket{X}$ and $\ket{x}$ correspond to the cluster assignments of different $q$'s, $(\sigma_{i}, \sigma_{j})$ with $q$ and $(\sigma^{\prime}_{i}, \sigma^{\prime}_{j})$ with $q^{\prime}$ ($q^{\prime} > q$), for a vertex pair $i$ and $j$. 
Because the labelling of the clusters is common to all of the vertices, we require that the refinement condition is satisfied with the common family of sets for every vertex pair. 
Under this condition, the KL divergence is nondecreasing as a function of $q$, which indicates that $E_{\mathrm{Bayes}}$ does not saturate earlier than $E_{\mathrm{Gibbs}}$. 
Similarly, $E_{\mathrm{training}}$ does not saturate earlier than $E_{\mathrm{Bayes}}$. 
Although the refinement condition that we required above is too strict to be satisfied exactly in a numerical calculation, it is what we naturally expect when the algorithm detects a hierarchical structure or the same structure with excess numbers of clusters. 
Moreover, the argument above is only a sufficient condition. 
Therefore, we naturally expect that $E_{\mathrm{Gibbs}}$ suggests a smaller number of clusters than $E_{\mathrm{Bayes}}$ and $E_{\mathrm{training}}$ quite commonly in practice. 
In addition, note that if we use a different criterion for the selection of $q$, e.g., the variation of the slope of the error function, the above conclusion can be violated.

\subsection*{Holdout method and $K$-fold cross-validation}
\begin{figure*}[t]
 \begin{center}
    \includegraphics[width=0.48 \columnwidth]{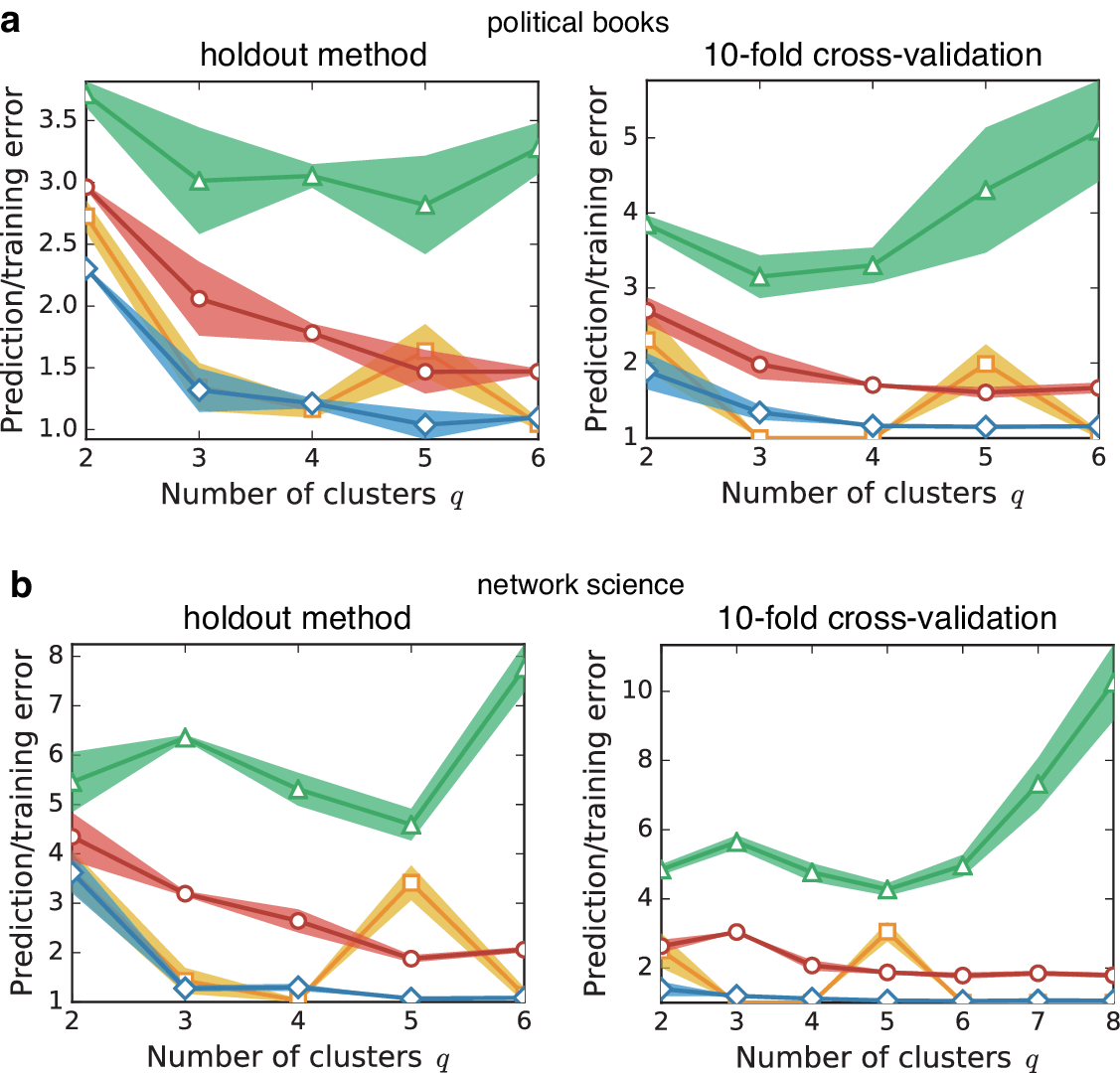}
 \end{center}
 \caption{
 	\textbf{Cross-validation errors using the holdout method and the $10$-fold cross-validation.} 
	As in Fig.~\ref{fig-politicalbooks}, the four data in each plot indicate 
	the Gibbs prediction errors $E_{\mathrm{Gibbs}}$ (green triangles), 
	MAP estimates $E_{\mathrm{MAP}}$ of $E_{\mathrm{Gibbs}}$ (yellow squares), 
	Bayes prediction errors $E_{\mathrm{Bayes}}$ (red circles), and 
	Gibbs training errors $E_{\mathrm{training}}$ (blue diamonds). 
	In the holdout method, we randomly selected $1\%$ of the edges ($5$ edges in (a) \textit{political books} and $10$ edges in (b) \textit{network science}) as the holdout set. The solid lines represent the average values obtained by repeating the prediction process $10$ times and the shadows represent their standard errors. 
	For the $K$-fold cross-validation, we set $K=10$. Although it is common to take average over different choices of $K$ partition, we omitted this averaging process. 
}
 \label{fig-holdout-Kfold}
\end{figure*}

Other than the LOOCV, it is also possible to measure the prediction errors by the holdout method and the $K$-fold cross-validation using BP. 
They can be conducted by randomly selecting the set of edges to be held out (i.e., the holdout set) and running BP that ignores the cavity biases sent by the held-out edges. 
In the holdout method, unlike the $K$-fold cross-validation, the test set is never used as the training set and vice versa. 
The results are listed in Fig.~\ref{fig-holdout-Kfold}. 
We observe that the overfitting of the Bayes prediction error $E_{\mathrm{Bayes}}$ appears to be prevented. 
Although the resulting behaviours are somewhat reasonable, we have to bear in mind that they have following conceptual and computational issues. 

First, we cannot expect the cross-validation assessment to work at all when the holdout set is too large, or equivalently, when the training set is too small. It is because, when a significant fraction of edges that are connected to vertices in certain clusters are held out, it is impossible to learn the underlying block structure correctly \cite{AiroldiNIPS2008}. 
Second, when the holdout set has more than one edge, we actually need to run the whole algorithm twice for each holdout set; first to compute the posterior distribution given the training set, and then, to evaluate the predictive distribution for the holdout set based on the obtained posterior distribution. 
Then, we need to repeat this process as many times as the number of holdout sets. 
Again, none of the above processes are required for the LOOCV, because we have the analytical expression. 
Thus, the computational cost for the holdout method and the $K$-fold cross-validation are orders of magnitude larger than the LOOCV. 

Besides the above issues, a more delicate treatment is required in the $K$-fold cross-validation. 
When the unobserved edges are connected, a prediction error is not given as an independent sum of the error per edge, and we need the calculation as we have done in equation (\ref{LeaveNodeOut2}). 
Therefore, the precise measurement is even more time consuming. 
In Fig.~\ref{fig-holdout-Kfold}, we imposed the constraint that the edges are not connected in the holdout method. 
For the $K$-fold cross-validation, we simply ignored the fact that the unobserved edges might be connected, although this discrepancy may not be negligible.

We lose all the advantages that we had in the LOOCV when the holdout method and the $K$-fold cross-validation are considered. 
However, when the computational efficiency can be compromised, the holdout method with small holdout set can be a promising alternative. 

\subsection*{Justification of the LOOCV estimates}
It is a common mistake that one will make use of the information of the test set (the unobserved edge in the present context) for the training of the model in the process of cross-validation. 
Precisely speaking, however, the information of the unobserved edge does contribute to the training of the model in our procedure; thus, the cross-validation estimate here is not correct in a rigorous sense. 
Nevertheless, the treatment that we propose is justified because the contribution of the unobserved edge in the present setting is negligible. 
First, when we update the distribution of the cluster assignment $\psi^{i \to j}_{\sigma}$, we make use of the information of the unobserved edge when there are loops in the network. 
Because we consider the sparse network, which is locally tree-like, the effect of the loops is typically negligible. 
Second, the update equations (\ref{SBMgammaUpdate}) and (\ref{SBMomegaUpdate}) of the parameters in the maximisation step of the EM algorithm contain the unobserved edge. 
However, because we consider the LOOCV in which only a single vertex pair is unobserved, this discrepancy results in a difference of only $O(N^{-1})$ to the parameter estimation. 
Therefore, our analytical results in terms of BP are sufficiently accurate to serve as error estimates when using the LOOCV.

\section*{Acknowledgments}
This work was supported by JSPS KAKENHI No. 26011023 (TK) and No. 25120013 (YK).

%\bibliographystyle{apsrev}
%\bibliographystyle{naturemag}
%\bibliography{bib-cavityBayesCV}

\end{document}